\documentclass[journal,comsoc]{IEEEtran}

\usepackage{booktabs} 
\usepackage{graphicx}
\usepackage{subcaption}
\usepackage{multirow,makecell}
\usepackage{mathtools}

\usepackage{tabularx}
\usepackage[linesnumbered,ruled,vlined]{algorithm2e}
\usepackage{balance}
\SetKwInOut{Input}{Input}
\SetKwInOut{Output}{Output}
\SetKwInOut{Require}{Require}

\usepackage{xcolor}
\usepackage{soul} 

\begin{document}

%

\title{ScalaBFS: A Scalable BFS Accelerator on HBM-Enhanced FPGAs}

\author{
	Kexin~Li,
	Chenhao~Liu,
	Zhiyuan~Shao,~\IEEEmembership{Member,~IEEE,}
	Zeke~Wang,~\IEEEmembership{Member,~IEEE,}
	Minkang Wu, 
	Jiajie Chen,
	Xiaofei~Liao,~\IEEEmembership{Member,~IEEE,} 
	and~Hai~Jin,~\IEEEmembership{Fellow ,~IEEE}
\thanks{K. Li, C. Liu, Z. Shao, M. Wu, J. Chen, X. Liao and H. Jin are with National Engineering Research Center for Big Data Technology and System, Services Computing Technology and System Lab, Cluster and Grid Computing Lab, School of Computer Science and Technology, Huazhong University of Science and Technology, Wuhan, 430074, P.R. China, E-mail:\{kx\_li, chenhao\_liu, zyshao, mk\_wu, bryce\_chen, xfliao, hjin\}@hust.edu.cn.}
\thanks{Z. Wang is with the Collaborative Innovation Center of Artificial Intelligence, Zhejiang University, P.R. China. E-mail: wangzeke@zju.edu.cn.}
\thanks{Corresponding author: Z. Shao, E-mail: zyshao@hust.edu.cn}
}

\IEEEtitleabstractindextext{
\begin{abstract}
High Bandwidth Memory (HBM) provides massive aggregated memory bandwidth by exposing multiple memory channels to the processing units. To achieve high performance, an accelerator built on top of an FPGA configured with HBM (i.e., FPGA-HBM platform) needs to scale its performance according to the available memory channels. In this paper, we propose an accelerator for BFS (Breadth-First Search) algorithm, named as ScalaBFS, that builds multiple processing elements to sufficiently exploit the high bandwidth of HBM to improve efficiency. 
We implement the prototype system of ScalaBFS and conduct BFS in both real-world and synthetic scale-free graphs on Xilinx Alveo U280 FPGA card (\emph{real hardware}). The experimental results show that ScalaBFS scales its performance almost linearly according to the available memory pseudo channels (PCs) from the HBM2 subsystem of U280. By fully using the 32 PCs and building 64 processing elements (PEs) on U280, ScalaBFS achieves a performance up to 19.7 GTEPS (Giga Traversed Edges Per Second). When conducting BFS in sparse real-world graphs, ScalaBFS achieves equivalent GTEPS to Gunrock running on the state-of-art Nvidia V100 GPU that features 64-PC HBM2 (twice memory bandwidth than U280).

\end{abstract}

\begin{IEEEkeywords}
FPGA; Hardware Accelerators; Breadth-First Search; Graph Analytics
\end{IEEEkeywords}}

\maketitle

\IEEEdisplaynontitleabstractindextext
\IEEEpeerreviewmaketitle

\section{Introduction}\label{sec:introduction}

BFS (Breadth-First Search) is one of the most fundamental algorithms of Graph theory and is widely used in application domains, such as navigation \cite{Ahuja:1990}, social network \cite{Bedi:2016}, and many others. The computation of BFS, however, is famous for its \emph{irregular memory accesses} and \emph{low computation-to-memory ratio} \cite{Umuroglu:2015}. To boost the performance of BFS, a computer system needs a large memory bandwidth, more specifically, a large memory bandwidth for random accesses. For such reason, BFS is chosen as one of the key benchmarks in Graph500 \cite{Richard:2010} to measure the capability of handling graph computing workloads of a computer system. 

For the low cost and massive storage capacity, DRAMs (i.e., DDR RAMs) are widely used as the storage devices in most computer systems nowadays. But DRAMs yield much lower bandwidth when handling random accesses than sequential accesses, and their single-channel performance is limited (e.g., 19.2GB/s for DDR4). To improve the memory bandwidth, newly emerging memory techniques, such as HMC (Hyper Memory Cube) \cite{Zhang:2017} and HBM (High Bandwidth Memory) \cite{Jun:2017}, are proposed. HMC stacks multiple DRAM dies and logic dies into a single stack and connects with computing units (e.g., CPU/GPU/FPGA) with a serial link providing bandwidth up to 240GB/s. Similarly, the HBM technique stacks multiple DRAM dies and a logic die into a single stack, but exposes multiple memory channels to the computing units. By this, HBM can easily scale its bandwidth with more stacks (thus more memory channels). For example, two stacks of HBM2 (the second generation of HBM) provide up to 460GB/s aggregated bandwidth in Xilinx U280 \cite{XilinxU280:2019}, while four stacks of HBM2 provide up to 900GB/s aggregated bandwidth in Nvidia V100 GPU \cite{NvidiaV100:2019}. 

The unprecedented huge memory bandwidth, especially the scaling memory channels provided by HBM, makes it possible to build efficient accelerators for bandwidth-critical workloads like BFS. Since with multiple memory channels, HBM can compensate the weakness on random memory accessing of the underlying DRAM devices. However, \emph{it is still challenging to design an FPGA-based BFS accelerator that sufficiently exploits the increasing number of memory channels, since the accelerator itself needs to be scalable in design to match HBM's massive memory bandwidth}.

To investigate this research direction, we design a BFS accelerator, named as ScalaBFS, in this paper. By attaching each memory channel with configurable amounts of processing elements, ScalaBFS can sufficiently exploits the memory bandwidth provided by an HBM to improve the performance of BFS. To the best of our knowledge, our work is the first system design that accelerates BFS on the FPGA-HBM platform.


Our paper makes the following contributions:

$\bullet$ proposes a design of BFS accelerator, that scales its performance according to the available memory channels of the FPGA-HBM platform. 


$\bullet$ provides an open-source \footnote{Available at: https://github.com/CGCL-codes/ScalaBFS} implementation that works on a real FPGA accelerator card (Xilinx Alveo U280) for data centers. 

$\bullet$ extensively evaluates our prototype system and compares its performance with state-of-art FPGA approaches as well as those on GPUs.

The rest of paper is organized as follows: Section \ref{sec:relatedworks} presents the background knowledge required to understand this work. Section \ref{sec:motivation} motivates the building of ScalaBFS. Section \ref{sec:systemoverview} elaborates the design of our accelerator. We discuss the performance scaling issues of ScalaBFS in Section \ref{sec:performancemodel}, and evaluate our prototype system in Section \ref{sec:performance}. Section \ref{sec:conclusion} concludes the paper and discusses the future works. 

\section{Background and Related Works}\label{sec:relatedworks}

In this section, we will first introduce the background knowledge on the BFS algorithm and HBM technology, and then briefly discuss the related works to our paper.

\subsection{BFS algorithm}\label{subsec:bfsalgorithm}

A directed \footnote{Undirected graphs can be easily converted into directed ones by treating each of its edges as two edges pointing to opposite directions.} graph $G = (V,E)$ consists of a set of a finite number of vertices $V$, and an edge set $E$ containing edges connecting two vertices from $V$. The BFS algorithm computes the distances of vertices in $V$ from a given $root$ vertex.\footnote{We omit the computing of \emph{parents} of the vertices in BFS algorithms, as it is trivial when compared with the computing of distance values.}

\begin{algorithm}[h]
	\setlength{\abovecaptionskip}{+5pt}
	\setlength{\belowcaptionskip}{-12pt}		
	\caption{BFS algorithm (push mode/pull mode).}
	\label{alg:bfs-general}
	\Input{Directed graph $G$, and \texttt{root} vertex $r$.}
	\Output{Array $Level[0...|V|-1]$, distances of vertices from $r$.}
	\DontPrintSemicolon
	\ForEach{$i\in [0,|V|-1]$}{
		$Level[i] \gets \infty$; 
	}
	$Level[r] \gets 0$; $bfs\_level \gets 0$; \\
	\tcp*[l]{push mode (beginning and ending iterations).}
	\While{$\exists i \in V$, such that $Level[i]==bfs\_level$ } {
		\ForEach{vertex $i$ whose $Level[i]==bfs\_level$}{
			\ForEach{outgoing neighbor $v$ of $i$}{
				\If{$Level[v] > bfs\_level$}{
					$Level[v] \gets bfs\_level + 1$;
				}
			}
		}
		$bfs\_level \gets bfs\_level + 1$; \\
	}
	
	\tcp*[l]{pull mode (mid-term iterations).}
	\While{$\exists i \in V$, such that $Level[i] == \infty$ } {
		\ForEach{vertex $i$ whose $Level[i] == \infty$}{
			\ForEach{incoming neighbor $u$ of $i$}{
				\If{$Level[u] == bfs\_level$}{
					$Level[i] \gets bfs\_level + 1$;
				}
			}
		}
		$bfs\_level \gets bfs\_level + 1$; \\
	}
\end{algorithm}

Our work considers the vertex-centric level synchronous implementation of BFS as listed in Algorithm \ref{alg:bfs-general}. The computing is organized into multiple iterations, where each while-loop in line 4$\sim$9 or 10$\sim$15 in the listing is treated as one iteration. Regarding to the direction of message passing, there are two modes of processing: push mode and pull mode. By push mode (line 4$\sim$9), an iteration browses the level array to locate all the \emph{active} (i.e., a vertex $i$, whose $Level[i]==bfs\_level$) vertices, and then for each of them, visits their outgoing (child) neighbors to set their level value to the current level if they are not visited before (i.e., pushing messages to the children). In pull mode (line 10$\sim$15), an iteration browses the level array to locate all the \emph{unvisited} vertices. Then for each of them, the algorithm visits their incoming (parent) neighbors, and sets its level value to the current level if one of its parents is active (i.e., pulling messages from the parents).

Generally, during the computation of the BFS algorithm, there will be small amounts of active vertices during the beginning and ending iterations, and large quantities of active vertices during the mid-term iterations \cite{Umuroglu:2015, Zhang:2018}. Therefore, choosing the push mode during the beginning and ending iterations, and switch to the pull mode during the mid-term iterations can help to reduce the number of memory access, and improve the performance of computation. 

For a scale-free graph (degrees of vertices comply with the power-law \cite{Faloutsos:1999}) containing millions of vertices, a push-mode (or pull-mode) iteration of BFS may contain hundreds of thousands of active (or unvisited) vertices. Although conflicts (e.g., different active vertices push level values to the same child vertex) exist, the algorithm logic of these active (or unvisited) vertices can be conducted in full parallel. The massively parallel nature of FPGAs makes them very suitable for the processing of the BFS algorithm. At the same time, the conflicts raised during computation can be comfortably solved using the on-chip memory resources (e.g., double pump BRAM used in our work) of the FPGAs. Consider the case that graph data are stored in the off-chip memory (e.g., HBM studied in this paper), which is regarded as the performance bottleneck of the system. The performance of BFS is then mainly decided by the time paid on reading the child (or parent) neighbors of the hundreds of thousands of active (or unvisited) vertices. \emph{With this regard, the throughput (not the latency) of the off-chip memory device plays a vital role in deciding the performance of BFS}. 

\subsection{High Bandwidth Memory (HBM)}\label{subsec:hbm}

\begin{figure}[t]
	\centering
	\setlength{\abovecaptionskip}{+5pt}
	\setlength{\belowcaptionskip}{-18pt}	
	\includegraphics[width = 3.2in]{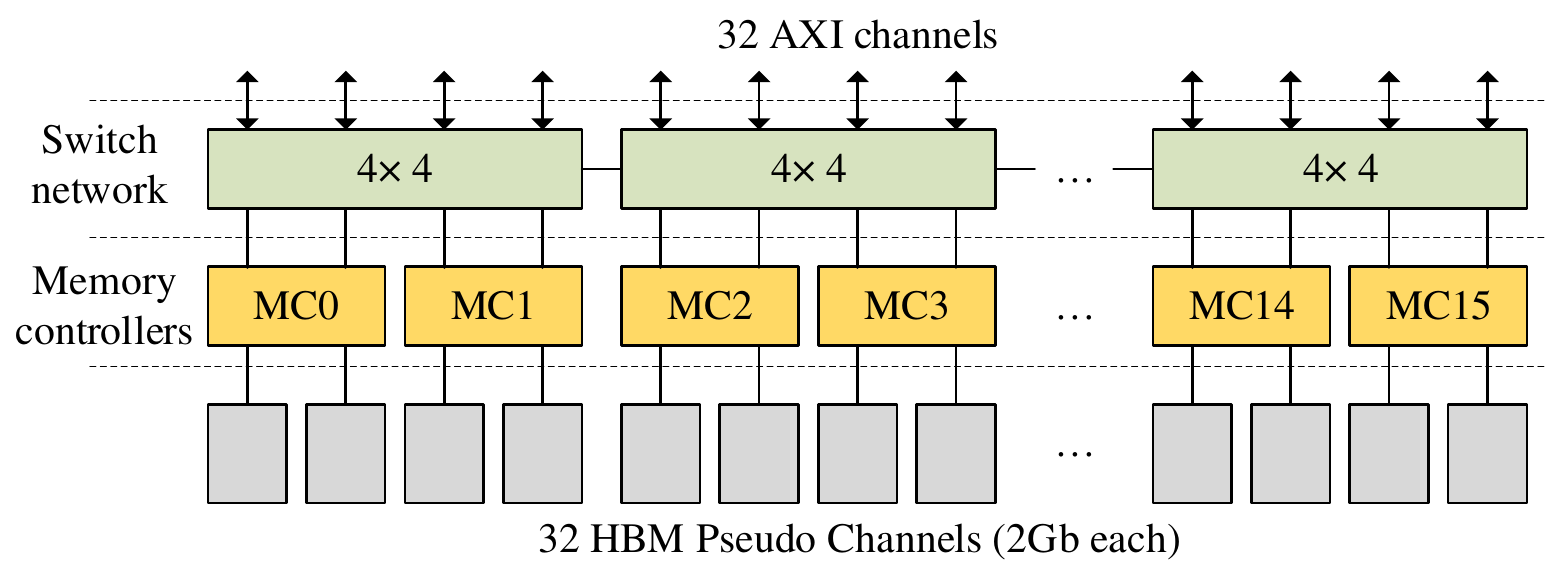}\	
	\caption{The HBM subsystem of Xilinx Alveo U280}
	\label{fig:u280hbm}
\end{figure}

As a typical example, Figure \ref{fig:u280hbm} illustrates the HBM subsystem of Xilinx Alveo U280 \cite{XilinxU280:2019}. The HBM subsystem contains two HBM stacks, each of which is divided into 16 pseudo channels (PCs). As each PC provides 2Gbit storage capacity, 32 HBM PCs provide 8GB ($2\times16\times2/8$) storage capacity totally.

Above the 32 PCs, there are 16 memory channels (MCs), each of which controls the accesses towards two adjacent PCs. The switch network consists of 8 4$\times$4 mini-switches, each of which connects two adjacent MCs and exposes 4 AXI interfaces to the FPGA. Thus, the HBM subsystem of U280 exposes 32 AXI ports totally to the user logic built in the FPGA. Every two adjacent mini-switches are connected with a bus to provide global addressing, such that a memory access towards an arbitrary PC can be issued from any AXI interface. 

By conducting random (but regular memory) accessing workloads, Shuhai \cite{Wang:2020} measured the performance of the HBM subsystem of U280, and observed that: 1) A memory access to HBM suffers higher latency than that towards DDR4. 2) Although the bandwidth of a single HBM PC is smaller than DDR4, the aggregated bandwidth of HBM arrives at 425 GB/s when handling sequential accesses. As BFS is in-essence a throughput-critical workload, the aggregated bandwidth of HBM can greatly help on boosting its performance, despite its longer latency for individual memory accesses.

\subsection{Graph data}
\begin{figure*}[t]
	\centering
	\begin{subfigure}{.2\textwidth}
		\centering
		\includegraphics[width=0.8\linewidth]{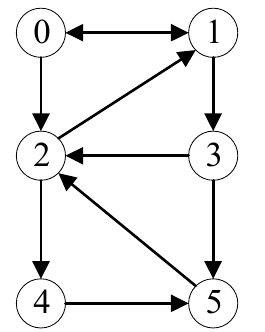}
		\caption{An example graph}
		\label{subfig:example}
	\end{subfigure}%
	\hfill
	\begin{subfigure}{.28\textwidth}
		\centering
		\includegraphics[width=1\linewidth]{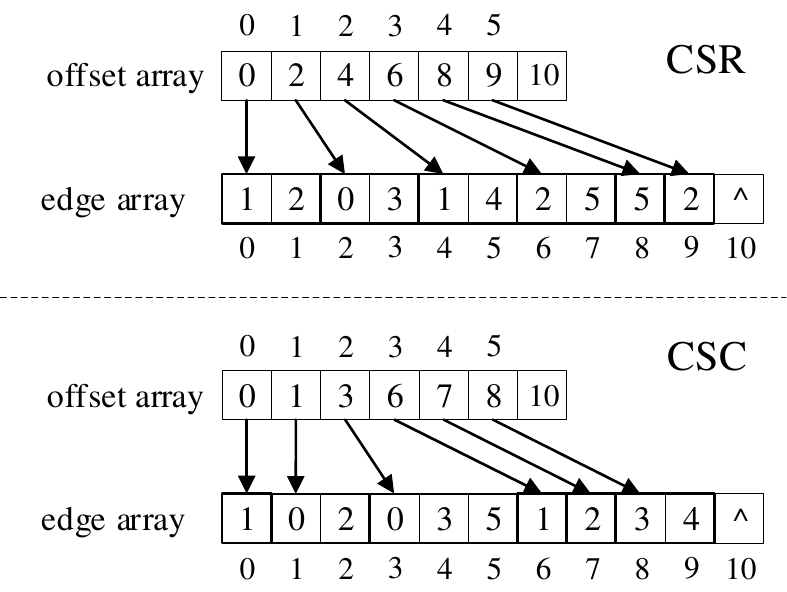}
		\caption{CSR \& CSC}
		\label{subfig:csrcsc}
	\end{subfigure}
	\hfill
	\begin{subfigure}{.28\textwidth}
		\centering
		\includegraphics[width=1\linewidth]{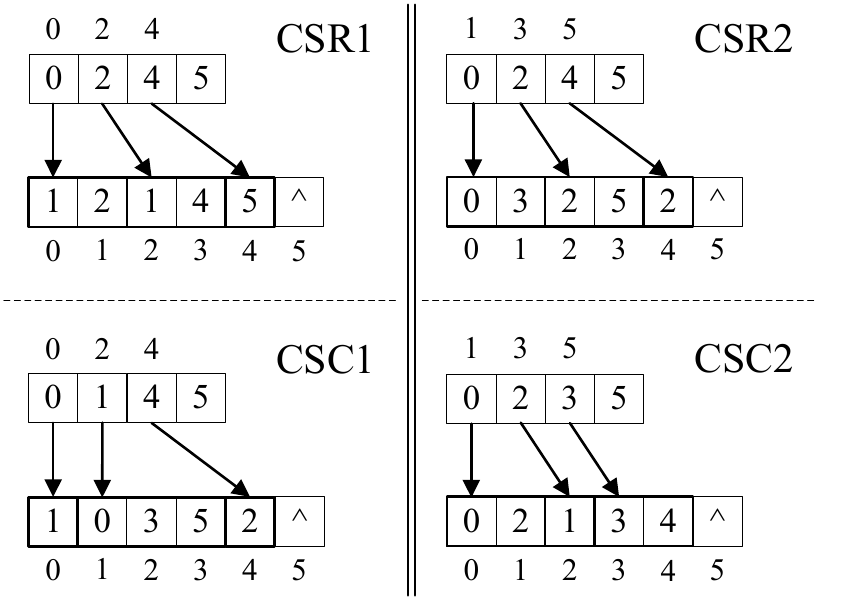}
		\caption{Divided into two subgraphs}
		\label{subfig:patitioned}
	\end{subfigure}
	\hfill
	\begin{subfigure}{.2\textwidth}
		\centering
		\includegraphics[width=1\linewidth]{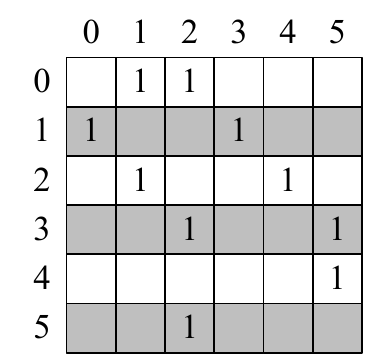}
		\caption{Adj. matrix view}
		\label{subfig:matrix}
	\end{subfigure}
	\caption{Partitioning an example graph into two subgraphs in CSC and CSR.}
	\label{fig:partitionexamplegraph}
\end{figure*}

We use Compressed Sparse Row (CSR) and its transpose, i.e., Compressed Sparse Column (CSC), to represent the graph data. The CSR consists of an offset array and an edge array, where the offset array stores the offsets of the outgoing (child) neighbor lists. The CSC has a similar structure as CSR, except that its edge array stores the incoming (parent) neighbor lists. The reason for ScalaBFS to use both CSR and CSC is that they facilitate the PEs to obtain both the outgoing and incoming neighbor lists of a selected vertex. Figure \ref{subfig:example} gives an example graph and its CSR and CSC representations in Figure \ref{subfig:csrcsc}.

\subsection{Related Works}\label{subsec:relatedworks}

As an important algorithm, the computation of BFS is extensively studied in various platforms, including CPUs \cite{Shun:2013, Zhang:2015}, GPUs \cite{Liu:2015, Wang:2017}, and FPGAs \cite{deLorimier:2006, Wang:2010, Betkaoui:2012, Attia:2014, Ni:2014, Lei:2015, Umuroglu:2015, Zhang:2017, Khoram:2018, Zhang:2018, Finnerty:2019}. We briefly survey the relevant works in FPGA-based BFS accelerators below. 

For a small graph that can be fully loaded in the on-chip memory of an FPGA, BFS can be conducted in it with very high performance using GraphStep \cite{deLorimier:2006}. However, for a large graph stored in the off-chip memory device (e.g., DRAM, HMC, or HBM studied in this paper), the key to improve the performance of BFS is to exploit the bandwidth provided by the off-chip memory devices, and at the same time, increase the parallelism of processing. With the advent of the big-data era, conducting BFS in the large graphs is of the research interest in most works. 

Works in \cite{Wang:2010} and \cite{Lei:2015} use bitmaps stored in the on-chip memory devices (e.g., BRAM) of FPGAs to track the status changing (active/unvisited) of vertices and focus on studying the interconnect networks connecting the processing elements. \cite{Wang:2010} proposes to use a fully-connected network and FIFOs to buffer the messages. The design requires $P^2$ FIFOs for $P$ processing elements. \cite{Lei:2015} constructs the network into a 2-D torus to reduce the resource consumption on FIFOs to $5\times P$. 

The work in \cite{Betkaoui:2012} and CyGraph \cite{Attia:2014} are two BFS accelerators build on the Convey HC-1 and HC-2 respectively. Convey HC-1 and HC-2 are customized machines \cite{Bakos:2010} whose coprocessor has a hardware 4$\times$8 full crossbar connecting 16 memory channels (8 memory controllers) with 4 Virtex-5 LX330 FPGAs. Both of these two accelerators repeatedly read the level array stored in off-chip DRAMs to track the status changing of vertices, and issue writes to the level array directly to the DRAMs (according to Algorithm \ref{alg:bfs-general}). By sufficiently using all memory channels and all logic resources of the 4 FPGAs, these two accelerators achieve the highest performance of 2.5 GTEPS on processing scale-free graphs. 

The work in \cite{Umuroglu:2015} proposes a hybrid BFS accelerator on the FPGA-CPU platform, where the CPU conducts the push-mode processing, and the FPGA executes the pull-mode processing. The work shows that choosing proper modes for different stages processing can effectively improve the performance of BFS. Recent work in \cite{Finnerty:2019} proposes an approach (named as Dr.BFS) that enhances the BFS performance by compressing the vertex data by using bitmaps. Dr.BFS can even achieve better performance than \cite{Umuroglu:2015}, with only push-mode processing. The methods on accelerating BFS on FPGA-HMC platform are studied in \cite{Zhang:2017}, \cite{Khoram:2018} and \cite{Zhang:2018}, where the main focuses are accelerating bitmap operations conducted in BFS with HMC memory devices. 

Besides the BFS accelerators studied in this paper, there are FPGA-based designs for general-purpose graph processing (i.e., supporting other graph algorithms, such as PageRank \cite{Brin:1998}, besides BFS), such as ForeGraph \cite{Dai:2017} and the work in \cite{Zhou:2016}. Aiming at supporting various graph algorithms, they generally adopt the edge-centric processing model to smooth the irregularity of graph processing workloads. The edge-centric processing model, however, limits their performances on BFS. For example, even after improvements on vertex caching \cite{Shao:2019}, ForeGraph achieves only about 410 MTEPS on the \texttt{soc-LiveJournal} graph (parameters listed in Table \ref{tab:graphs}) according to the metrics of Graph500 by using a single DDR4 channel. 

\section{Opportunity and Challenge}\label{sec:motivation}

In synchronous vertex-centric BFS algorithm~\ref{subsec:bfsalgorithm}, the vertices of the current level can be processed in parallel. However, the neighbor update of every vertex can have overlaps, thus need to be combined before written into level array or visited map. For example, the state-of-the-art implementation Dr. BFS~\cite{Finnerty:2019} builds a status recombiner to combine those updates after every iteration. When the parallelism increases, the recombiner will need a longer time to process the updates due to limited external memory bandwidth, which becomes the performance bottleneck of the overall system~\cite{Finnerty:2019}. The underlying reason is that previous FPGA boards always feature up to two DDR4 channels, providing up to 40GB/s memory bandwidth. Therefore, exploiting more computing parallelism fails to increase the overall performance and then the FPGA resources are underutilized even in a small FPGA~\cite{Finnerty:2019}. For example, Dr. BFS~\cite{Finnerty:2019} utilizes 27\% of the FPGA logic and Fabgraph~\cite{Shao:2019} uses only 19\%.

\subsection{Opportunity from HBM-enhanced FPGAs}
FPGA vendors such as Xilinx have already provided HBM-enhanced FPGA boards like U280 to provide up to 460GB/s memory bandwidth~\cite{Wang:2020}, which opens up new opportunity to accelerate graph processing in the context of FPGA. Therefore, HBM-enhanced FPGAs allow more computing parallelism to fully leverage FPGA resources. However, it is not trivial to efficiently utilize HBM bandwidth in the context of BFS, which introduces random memory access.   

\subsection{Challenge when Leveraging HBM}
Due to the random memory access property of BFS, we need a crossbar to globally access all the memory channels of HBM. There are two types of crossbar, which turns out to fail to satisfy the memory bandwidth requirement of our BFS engine. 

\noindent{\bf A Full Xilinx AXI Interconnect. }In our practice, the full Xilinx AXI Interconnect IP \cite{XilinxAXIInterconnect:2017} only supports up to 16$\times$16 AXI interconnect, which is unacceptable as U280 has 32 HBM PCs. 

\begin{figure}[h]
	\centering
	\includegraphics[width=8.2cm]{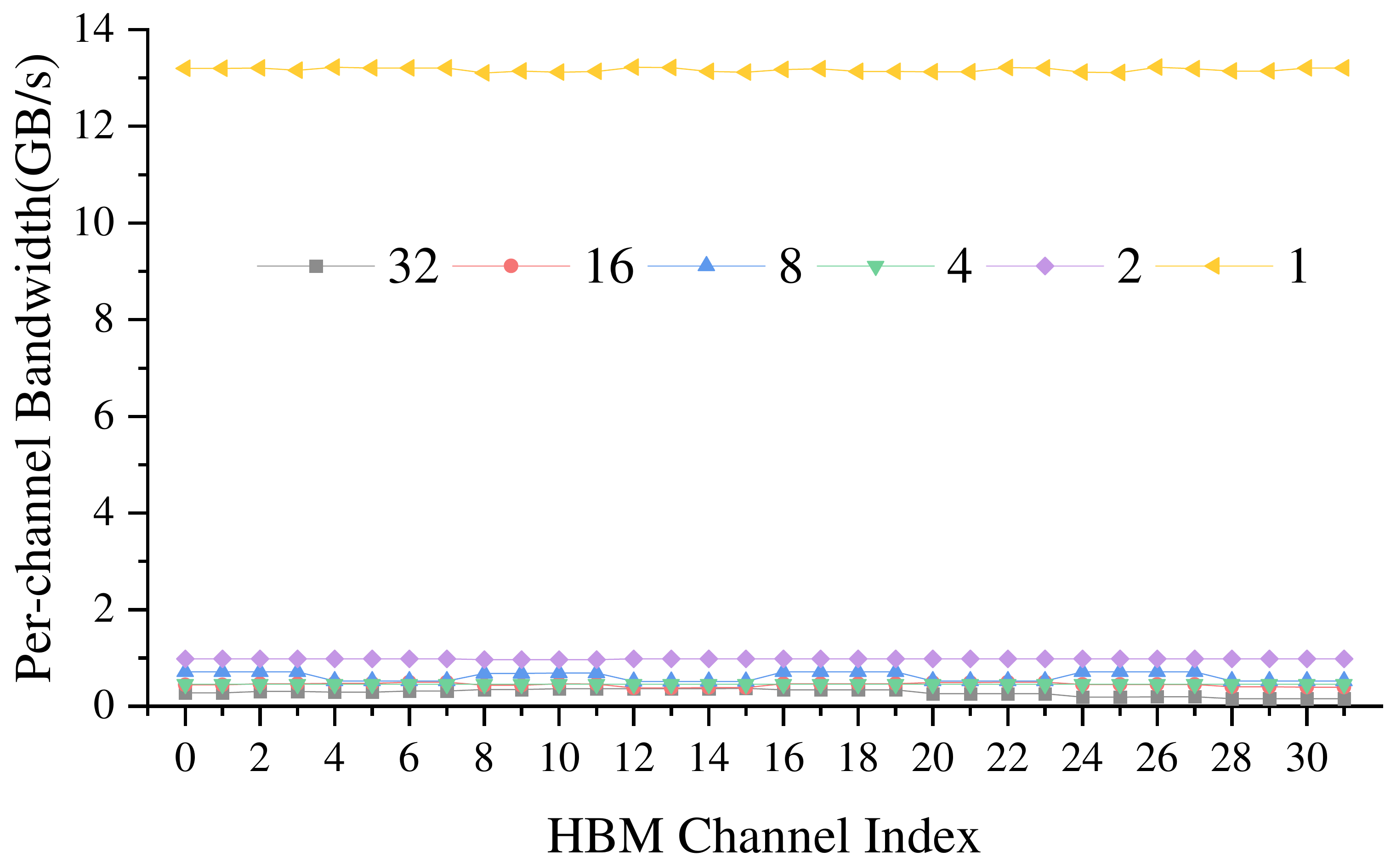}
	\vspace{-1ex}
	\caption{Effect of switch network when the application reads cross HBM channels. }
	\vspace{-1ex}
	\label{fig_effect_mini_switch}
\end{figure}

\noindent{\bf Xilinx's Switch Network. }We examine the effect of the built-in Xilinx's switch network when memory accesses cross HBM channels. We employ Shuhai~\cite{Wang:2020} to benchmark the performance of each AXI channel by reading data from neighboring HBM channels. Our data width is 256 bits, with the outstanding buffer size 256, and the burst length 64. Intuitively, the more number of HBM channels each AXI channel reads from, the higher pressure the switch network encounters. Figure~\ref{fig_effect_mini_switch} illustrates the throughput of each AXI channel that reads across $2^k$ neighboring HBM channels, where $k=0,1,...,5$. 
For example, "4" in the figure legend indicates that the $i^{th}$ $(0 \leq i \leq 31)$ AXI channel in Shuhai visits 4 consecutive HBM PCs in the range of [$\lfloor i/4\rfloor \cdot4$, $\lfloor i/4\rfloor \cdot4+3$] in every 4 memory accesses.

We observe that reading across HBM channels significantly affects the achievable throughput of each AXI channel. For example, the case that crosses 32 HBM channels achieves less than 0.5GB/s, more than 20 times less than the case that does not cross any HBM channels. It means that globally random memory access leads to dramatically low throughput using Xilinx's built-in switch network. 


\section{System Design}\label{sec:systemoverview}

In this section, we present the system design of ScalaBFS. We present the design methodology behind and the overall hardware architecture of ScalaBFS, followed by detailed hardware design of each hardware component.

\subsection{Design Methodology}\label{subsec:design_methodology}
In this subsection, we present two design goals of ScalaBFS, followed by our solution. 

\subsubsection{Two Design Goals of ScalaBFS} 
$\\$
\noindent{\bf G1: Minimize Off-chip Memory Accesses Given Limited On-chip Memory Capacity. }The size of on-chip memory, i.e., BRAMs and URAMs, is always limited, especially on FPGAs, where memory blocks are uniformed distributed. For example, the RAM size on state-of-the-art FPGA board U280 is 41MB~\cite{XilinxU280:2019}. In the context of BFS, we are not able to fit all the data (vertex data and graph data) in FPGA's on-chip memory, which means we have to read data from external memory. Since external memory access is quite expensive, we aim at reducing external memory accesses as many as possible. 

\noindent{\bf G2: Maximize Computing Parallelism to Match High HBM Bandwidth. }Previous work~\cite{Finnerty:2019} can only achieve limited computing parallelism due to limited external memory bandwidth. In other words, it cannot fully utilize FPGA resources on a small-size FPGA as external memory bandwidth has already been saturated. On an HBM-enhanced FPGA, we are able to explore high computing parallelism to match high HBM bandwidth. However, it is not trivial to achieve high computing parallelism, because the parallel implementation of BFS is not able to guarantee execution locality, so we need data shuffling that increases design complexities and impedes us from easily achieving linear scalability. 



\subsubsection{Our Solution}
$\\$

In order to satisfy {\bf G1}, we need to minimize 1) reference input graph from external memory {\bf M1}, and 2) the on-chip memory overhead for intermediate states {\bf M2}. 

\noindent{\bf Satisfying M1. }For an input graph, we separate its data into two types: vertex data and graph data. Vertex data refers to the level array, while graph data refers to the neighbor lists. We leverage precious on-chip memory resources (i.e., BRAMs and URAMs) to store all the vertex data, which is similar to \cite{Wang:2010} and \cite{Lei:2015}. The underlying reason behind this design choice is two-fold. 
First, Algorithm \ref{alg:bfs-general} illustrates that vertex data is frequently modified to track the status (i.e., active or visited) of vertices, and at the same time, store the result level values. On the contrary, graph data never change. 
Second, modern FPGAs now enjoy larger and larger on-chip memory capacity, in terms of BRAM and URAM, as FPGA technique advances. Larger on-chip memory capacity allows storing millions of vertices.\footnote{Discussing the case the size of vertex data is larger than the on-chip memory capacity is beyond the scope of this paper. }

\noindent{\bf Satisfying M2. }We present a new BFS algorithm to minimize intermediate state size, as listed in Algorithm \ref{alg:bfs-optimize}. The key idea of the proposed algorithm is to employ three bitmaps, i.e., \emph{current frontier}, \emph{next frontier} and \emph{visited map}, to track the statuses of vertex data during the execution of BFS. Each vertex occupies a single bit from each of these three bitmaps, which means that a vertex only consumes three bits. A bit in the current frontier indicates whether its corresponding vertex is active (1 for active, and 0 for inactive) in the current iteration. Similarly, a bit in the next frontier indicates whether its corresponding vertex will be activated in the next iteration. A bit in the visited map indicate whether the corresponding vertex is visited before (1 for visited, and 0 for un-visited). These three bitmaps are stored in double pump BRAMs, such that two operations can be conducted on them within a clock cycle. 

\begin{algorithm}[h]
	\caption{BFS algorithm using three bitmaps}
	\label{alg:bfs-optimize}
 	\setlength{\abovecaptionskip}{+5pt}
	\setlength{\belowcaptionskip}{-15pt}	
	\Input{Directed graph $G$, and \texttt{root} vertex $r$.}
	\Output{Array $Level[0...|V|-1]$, distances of vertices from $r$.}
	\DontPrintSemicolon
	\ForEach{$i\in [0,|V|-1]$}{
		$Level[i] \gets \infty$; $current\_frontier[i] \gets 0$; \\
		$next\_frontier[i] \gets 0$; $visited\_map[i] \gets 0$; \\
	}
	$Level[r] \gets 0$; $bfs\_level \gets 0$;\\
	$current\_frontier[r] \gets 1$; $visited\_map[r] \gets 1$;\\
	
	\tcp*[l]{push mode (beginning and ending iterations).}
	\While{$\exists i \in V$, such that $current\_frontier[i]=1$ } {
		\ForEach{vertex $i$ whose $current\_frontier[i]=1$}{
			\ForEach{outgoing neighbour $v$ of $i$}{
				\If{$visited\_map[v] == 0$}{
					$next\_frontier[v] \gets 1$;\\
					$visited\_map[v] \gets 1$;\\
					$Level[v] \gets bfs\_level + 1$;
				}
			}
		}
		$bfs\_level \gets bfs\_level + 1$; \\
		$swap(current\_frontier, next\_frontier)$
	}
	
	\tcp*[l]{pull mode (mid-term iterations).}
	\While{$\exists i \in V$, such that $visited\_map[i]\neq1$ } {
		\ForEach{vertex $i$ whose $visited\_map[i]\neq1$}{
			\ForEach{incoming neighbour $u$ of $i$}{
				\If{$current\_frontier[v] == 1$}{
					$next\_frontier[i] \gets 1$;\\
					$visited\_map[i] \gets 1$;\\
					$Level[i] \gets bfs\_level + 1$;
				}
			}
		}
		$bfs\_level \gets bfs\_level + 1$; \\
		$swap(current\_frontier, next\_frontier)$
	}
\end{algorithm}

In order to satisfy {\bf G2}, ScalaBFS partitions the graph data into multiple subgraphs, and places each subgraph in an HBM PC to enforce locality of accessing (prevent the crossing shown in Figure \ref{fig_effect_mini_switch}). Figure \ref{fig:partitionexamplegraph} illustrates how to divide the vertex ID space of an example graph into two PEs: first, IDs are divided into two intervals, i.e., [0,2,4] and [1,3,5], due to the load-balancing reason. Then, the graph data are partitioned according to the partitioning results on the vertex ID space: neighbor lists of the vertices belonging to the same partition will be placed in the same subgraph as shown in Figure \ref{subfig:patitioned}. From the viewpoint of the adjacency matrix, our partitioning scheme partitions a graph \emph{horizontally}, which is different from \cite{Finnerty:2019} where the input graph is partitioned vertically. The reason for this horizontal partition scheme is that it does not breakdown the neighbor lists, and longer neighbor lists mean more chances of sequential accesses towards the HBM, which helps on improving the memory bandwidth usage rates during processing. 

Further, ScalaBFS employs multiple computing engines to provide massive computing parallelism, and the details are shown in the following hardware design.

\begin{figure}[t]
	\centering
 	\setlength{\abovecaptionskip}{+5pt}
 	\setlength{\belowcaptionskip}{-15pt}
	\includegraphics[width = 3in]{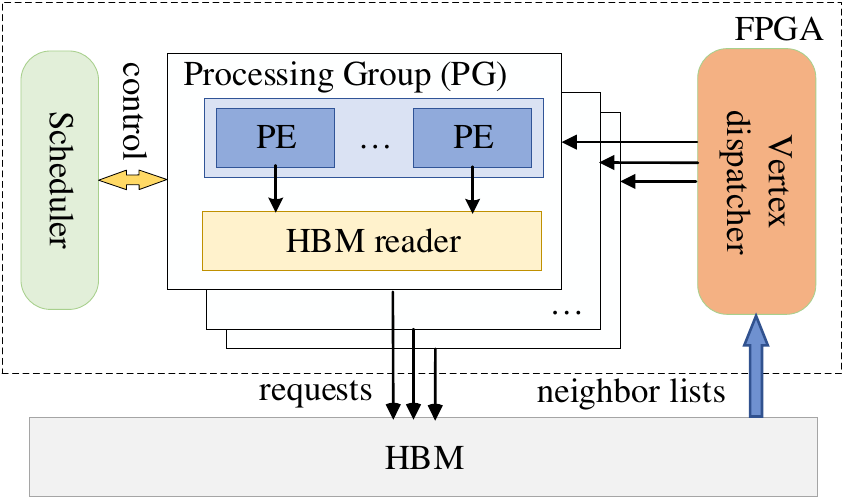}	
	\caption{Architecture of ScalaBFS}
	\label{fig:architecture}
\end{figure}

\begin{figure*}[ht]
	\centering
	\setlength{\belowcaptionskip}{-10pt}	
	\begin{subfigure}{.45\textwidth}
		\centering
		\includegraphics[width=1\linewidth]{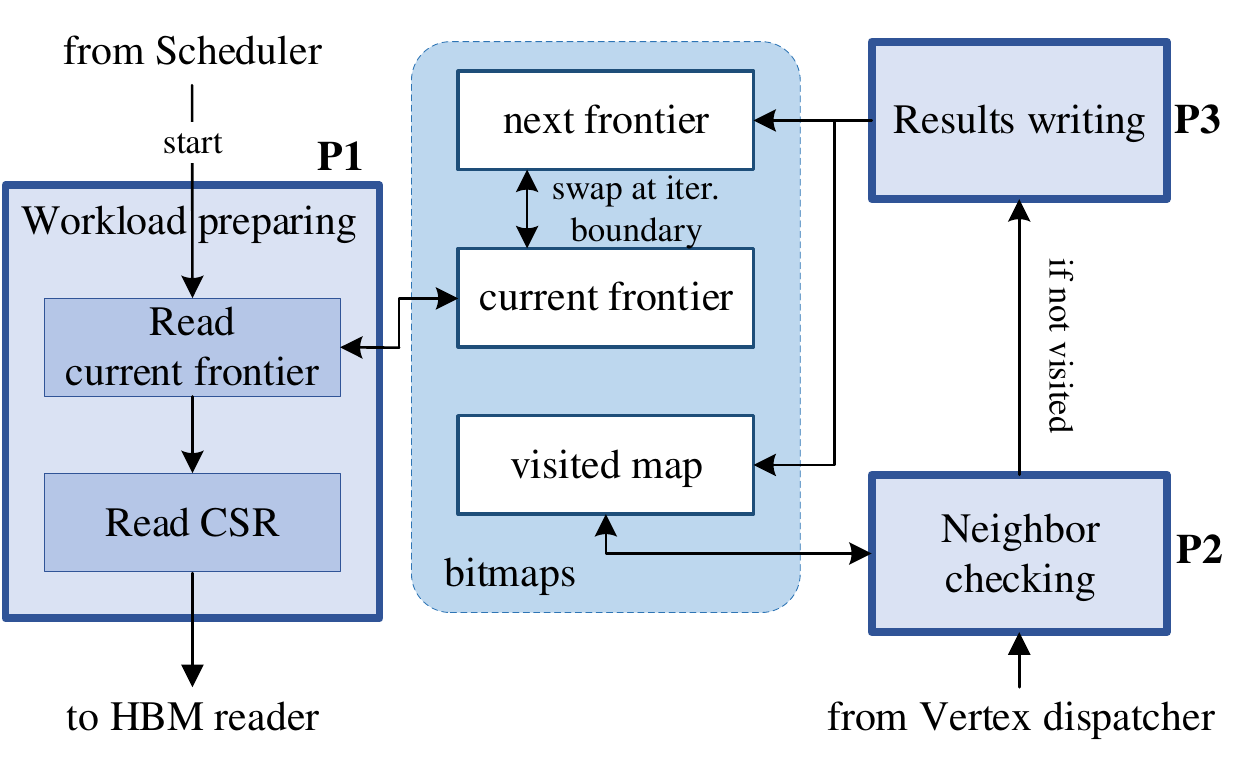}
		\caption{Push mode}
		\label{subfig:pushmode}
	\end{subfigure}
	\hfill
	\begin{subfigure}{.45\textwidth}
		\centering
		\includegraphics[width=1\linewidth]{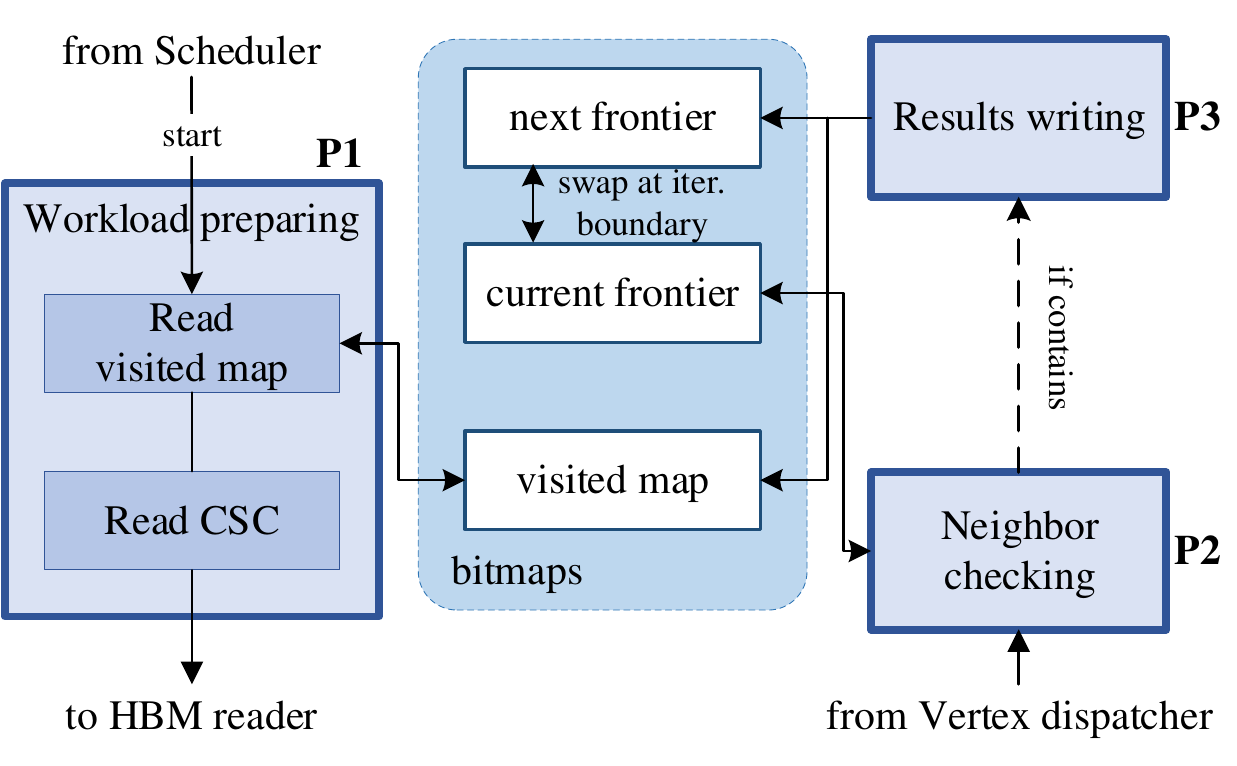}
		\caption{Pull mode}
		\label{subfig:pullmode}
	\end{subfigure}
	\caption{Processing logic of a PE in both push and pull modes.}
	\label{fig:hybridpipelines}
\end{figure*}

\subsection{Overall Hardware Architecture}
\label{subsec:overall_architecture}

According to Algorithm~\ref{alg:bfs-optimize}, we present the hardware architecture of ScalaBFS, which consists of multiple \emph{Processing Groups} (PGs), a \emph{Scheduler}, and a \emph{Vertex dispatcher}. A PG consists of an \emph{HBM reader} and one or more \emph{Processing Elements} (PEs). The Scheduler in Figure \ref{fig:architecture} controls the processing mode (either pull or push model) of each PE, and informs its decisions to the PEs at the beginning of each iteration on the fly. 

Each PG is assigned exclusively to a single HBM pseudo channel (PC). The HBM reader is shared by all the PEs in a PG to issue memory requests to its corresponding HBM PC via AXI port to read the neighbor lists from HBM. Therefore, ScalaBFS has no more than 32 PGs on Xilinx Alveo U280, as its HBM subsystem has only 32 PCs (shown in Figure \ref{fig:u280hbm}). 

Each PE processes an \emph{interval} of vertices of an input graph. In case of total $Q$ PEs, we divide the vertex ID space of an input graph into $Q$ non-overlapping intervals. In order to achieve load balancing between PEs, the vertex IDs are hashed before assigning to the intervals, such that the $i$-th PE is in charge of processing the vertex whose $VID$ satisfies $VID \% Q = i$. Moreover, Each PE works in a hybrid (push-pull) processing mode according to the stages of BFS, to improve hardware utilization rate. The design details of a PE is shown in Subsection~\ref{subsec:hybridpe}.

After HBM readers issue memory requests, the Vertex dispatcher gathers all responses of neighbor lists from all the PCs and then scatters the vertices in the neighbor lists to the destination PEs according to their vertex IDs (denoted as $VID$). We will present the details of these two modules in Subsection~\ref{subsec:readeranddispatcher}.

\begin{figure*}[ht]
	\centering
	\setlength{\belowcaptionskip}{-10pt}	
	\begin{subfigure}{.28\textwidth}
		\centering
		\includegraphics[width=0.9\linewidth]{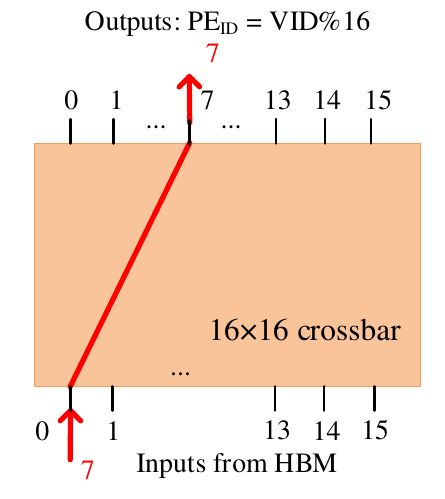}
		\caption{full crossbar}
		\label{subfig:fullcrossbar}
	\end{subfigure}%
	\hfill
	\begin{subfigure}{.65\textwidth}
		\centering
		\includegraphics[width=0.9\linewidth]{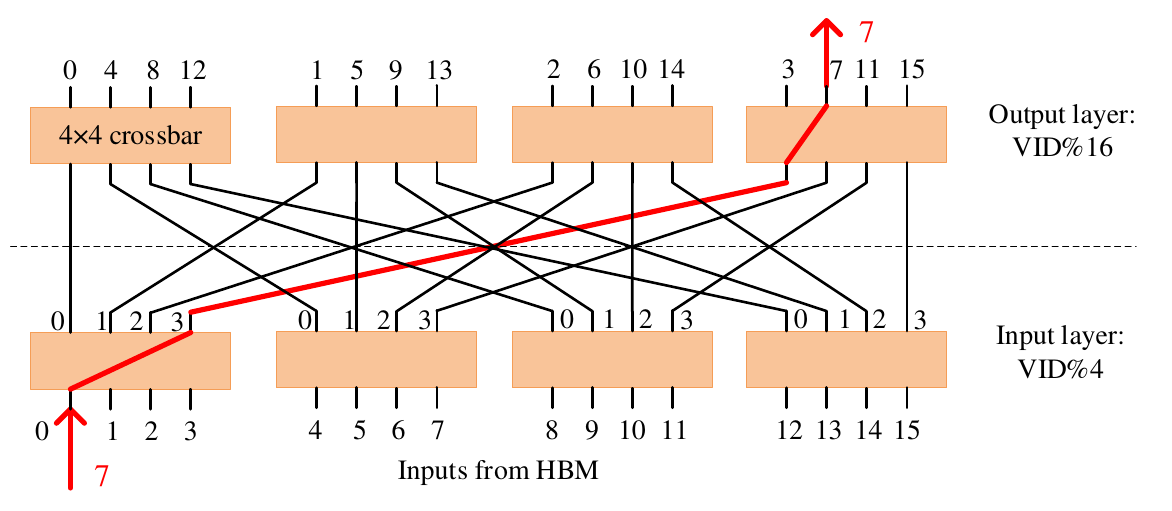}
		\caption{equivalent 2-layer crossbar}
		\label{subfig:2layercrossbar}
	\end{subfigure}
	\caption{Converting full-crossbar into multi-layer crossbar.}
	\label{fig:crossbar}
\end{figure*}

\subsection{Hybrid-Mode PE}
\label{subsec:hybridpe}
The goal of a hybrid-model PE is to allow both push and pull models within a PE. Intuitively, a PE works in the push mode in the beginning and ending iterations, and work in the pull mode during the mid-term iterations as listed in Algorithm \ref{alg:bfs-optimize}. A hybrid-model PE consists of on-chip intermediate states and three execution components (i.e., stages).

\noindent\textbf{Working Mechanism of Intermediate States. }During a push-mode iteration, the current frontier is scanned to locate all the active vertices. And then for each active vertex, vertices in its outgoing (child) neighbor list will be checked. If a vertex in the list has not been visited before (by checking the visited map), the corresponding bit in the visited map and next frontier will be set, and its level value will be written back to the level array stored in URAMs. Whereas during a pull-mode iteration, the visited map will be scanned to locate all unvisited vertices. And then for each unvisited vertex, vertices in its incoming (parent) neighbor list will be checked. If a parent vertex in the list is active (by checking the current frontier), the corresponding bits in the visited map and next frontier of the aforementioned unvisited vertex will be set, and its level value will be written back to URAMs. 



\noindent\textbf{Workload Preparing (P1). }
This stage prepares the workloads to be processed. When working in the push mode, the active vertices are first obtained from the current frontier, and then passed to the \emph{Read CSR} module to read their outgoing neighbor lists from the CSR data. As the CSR data are stored in HBM, the module will pass its requests to the HBM reader. When working in the pull mode, the unvisited vertices are obtained from the visited map, and the remaining processes are the same as in the push mode except that the \emph{Read CSC} module will read the CSC data for incoming neighbor lists from the HBM. 

\noindent\textbf{Neighbor Checking (P2). }
This stage accepts messages (i.e., the neighboring vertices) from the Vertex dispatcher. When working in the push mode, the messages will be the outgoing (child) neighbors of the active vertices prepared in P1. For each child vertex, the processing logic will proceed to check the visited map, and pass it to the next stage of processing (i.e., P3) if it is not visited before, or drop it if otherwise. When working in the pull mode, the messages will be the incoming (parent) neighbors of the unvisited vertices prepared by the stage of P1. For each parent vertex, the processing logic will then proceed to check the current frontier, and pass \emph{its child vertex} (i.e., the unvisited vertex prepared by P1) to the next stage of processing (i.e., P3) if it (the parent vertex) is active, or drop it if otherwise. 

Note that in the pull mode, the child vertex to be passed to P3 may \emph{not} be in the same PE as that processing its parent, since the incoming neighbor lists in CSC contains vertices of other partitions as shown in Figure \ref{subfig:patitioned}. In such a case, the child vertex will be passed from one PE to another PE via a soft crossbar, which will be discussed in detail in the next subsection. 

\noindent\textbf{Result Writing (P3). }
This stage accepts result messages from P2 and conducts the modifications in the bitmaps according to Algorithm \ref{alg:bfs-optimize}. In both processing modes, the results will be written to the next frontier and visited map. Besides, the results will also be written to the Level array stored in URAMs (not shown in Figure \ref{fig:hybridpipelines} for brevity). 

Due to the similarities of the logic for these two processing modes, except for the soft crossbar used in P2 in the pull mode, we use the same sets of circuits with parameters in registers to implement the similar modules, to save logic resources of the FPGA. Moreover, as the above processing stages are driven by signals (from the Scheduler) or data streams from other modules, the circuits work asynchronously in a pipelined fashion to maximize processing efficiency.

\subsection{HBM Reader and Vertex Dispatcher}
\label{subsec:readeranddispatcher}

\noindent{\bf HBM Reader. }The functionality of an HBM reader is to receive requests from the \emph{Read CSR} or \emph{Read CSC}, convert these requests into memory accessing AXI commands, and then issue them to its corresponding HBM PC. When PEs are now working in the push mode. A request (reading the outgoing neighbor list of an active vertex) is sent to an HBM reader, which will first assemble an AXI command to read the corresponding values in the offset array of CSR from HBM. After receiving the offset values, the HBM reader will then assemble another AXI command to read the outgoing neighbor list from the edge array. Procedures are similar in the pull mode, except that each PE will read the offsets in the CSC and incoming neighbor lists. 

\noindent{\bf Vertex Dispatcher. }The Vertex dispatcher \emph{intercepted} read the neighbor lists read from the HBM. particularly, The functionality of the Vertex dispatcher is to scrutinize the vertices in the input neighbor-list streams, classify them according to intervals that they belong to, and send them back to the corresponding PEs. The most straightforward approach that achieves such an objective is to use a full $N\times N$ crossbar where N is the number of PEs (and accordingly, $N$ subgraphs). However, such a full crossbar requires $N^2$ FIFOs to implement \cite{Wang:2010}, which are hard to fit within an FPGA when $N$ is sufficiently large. For example, when the length of FIFO is 16 and $N=64$, the Vertex dispatch employs a full 64$\times$64 crossbar that consumes more than half of the LUTs in the U280 FPGA card, leaving very limiting number of LUTs for PEs.  

\noindent{\bf Efficient Crossbar Design. }Inspired by the fact that the switching logic of our Vertex dispatcher is unidirectional (and thus simpler), we propose a \emph{multi-layer crossbar} that requires much fewer FPGA resources to implement while keeping the same functionality. \noindent{\bf Efficient Crossbar Design. }Inspired by the fact that the switching logic of our Vertex dispatcher is unidirectional (and thus simpler), we propose a \emph{multi-layer crossbar} that requires much fewer FPGA resources to implement while keeping the same functionality. Our multi-layer crossbar design in ScalaBFS shares some similarities as the recent work of butterfly crossbar in \cite{choi:2021}. The differences between our design and that in \cite{choi:2021} are that: 1) our crossbar design supports general RTL (as ScalaBFS), while that in \cite{choi:2021} mainly supports HLS (High Level Synthesis); 2) the focus of our crossbar design is to cope with random and irregular memory accesses, while that in \cite{choi:2021} mainly addresses sequential memory accesses; 3) our crossbar design studies the trade-off between performance and on-chip resource consumption (explained in the following), while there is no such concern in \cite{choi:2021}. 

To explain our multi-layer crossbar, we discuss a case that dispatches 16 neighbor-list streams to 16 PEs as shown in Figure \ref{fig:crossbar}. Figure \ref{subfig:fullcrossbar} depicts the 16$\times$16 full crossbar that contains 16 input and 16 output ports. The switching logic of this 16$\times$16 full crossbar is to send a vertex to the $(VID\%16)^{th}$ port, where $VID$ denotes the vertex's ID. 
Equivalently, we can convert the 16$\times$16 full crossbar into a two-layer crossbar, which consists of two layers (input layer and output layer) and each layer consists of four 4$\times$4 crossbar, as shown in Figure \ref{subfig:2layercrossbar}. The switching logic of each 4$\times$4 crossbar of the input layer is to send a vertex to its $(VID\%4)^{th}$ port, which connects to the $(VID\%4)^{th}$ 4$\times$4 crossbar of the output layer. The $i^{th}$ 4$\times$4 crossbar of the output layer connects to the PEs such that $PE_{ID}\%4 = i$, where $PE_{ID}$ denotes the ID of a PE. By this for the output layer, the first ($0^{th}$) 4$\times$4 crossbar connects to $PE_0$, $PE_4$, $PE_8$ and $PE_{12}$, while the second ($1^{th}$) 4$\times$4 crossbar connects to $PE_1$, $PE_5$, $PE_9$ and $PE_{13}$, and so on. The switching logic of each 4$\times$4 crossbar of the output layer is thus to send an incoming vertex to the $(VID\%16)^{th}$ PE (equals to the effectiveness of the 16$\times$16 full crossbar). 

To generalize our approach on converting an $N\times N$ full crossbar into a multi-layer crossbar: we need first decompose $N$ into multiple (say $k$) factors, such that $N = C_1\times C_2 \times ... \times C_k$. The first (input) layer uses $N/C_1$ $C_1\times C_1$ crossbars, and classifies the input vertices into $C1$ groups according to $VID\%C_1$, where $VID$ denotes the ID of an input vertex. The second (relay) layer uses $N/C_2$ $C_2\times C_2$ crossbars, and further classifies the input vertices into $C1\times C2$ groups according to $VID\%(C_1 \times C2)$. Such a process continues till the vertices arrive at the last (output) layer, which uses $N/C_k$ $C_k\times C_k$ crossbars, and classifies the vertices into $C_1\times C_2\times ... \times C_k = N$ groups. The ($N$) vertex groups coming out of the last layer are finally sent to the ($N$) PEs, each of which is in charge of processing one vertex group (i.e., a vertex interval). Figure \ref{fig:crossbar} merely illustrates a simple case where $N=16$, $k=2$, and $C_1=C_2=4$. 

We now compare these two approaches (full-crossbar vs. multi-layer crossbar) from the angles of resource consumption and efficiency. An $N\times N$ full crossbar consumes $N^2$ FIFOs, while the number of FIFOs required by its equivalent $k$-layer crossbar is $(N/C_1) \times C_1^2 + (N/C_2) \times C_2^2 + ... + (N/C_k) \times C_k^2$. It is easy to prove that the number of FIFOs consumed by our $k$-layer crossbar is smaller than that of its equivalent $N\times N$ full crossbar, as $N = C_1\times C_2 \times ... \times C_k$. Consider the example in Figure \ref{fig:crossbar}, the 16$\times$16 full crossbar consumes $16\times 16 = 256$ FIFOs, while the two-layer crossbar consume only $2\times 4\times 4\times4 = 128$ FIFOs, meaning half resource consumption. 

From the efficiency point of view, the $N\times N$ full-crossbar achieves 1-hop latency on message passing, while the equivalent $k$-layer crossbar approach requires k-hop latency for vertex dispatching. Obviously, the $k$-layer crossbar approach leads to higher latency than the full crossbar approach as $k\geq 2$. Nevertheless, based on the observation that BFS is a throughput-critical workload as discussed in subsection \ref{subsec:bfsalgorithm}, it is appropriate for us to use multi-layer crossbars to trade latency for resource. In Section \ref{sec:performance}, we will use 3-layer 4$\times$4 crossbars to replace the 64$\times$64 full crossbar to help our accelerator to achieve the scale of 64 PEs on U280. 

\section{Performance Model} \label{sec:performancemodel}

ScalaBFS can scale its performance in two directions: 1) increasing the number of HBM PCs (thus PGs) with a fixed number of PEs in a PG, and 2) increasing the number of PEs in a PG with a fixed number of HBM PCs. The first direction of scaling is limited by the number of available HBM PCs of a given HBM subsystem, while scaling in the second direction may be limited by the amount of logic resources of an FPGA. If we regard the HBM as the slower device, and the processing circuits (PEs, HBM readers, and the Vertex dispatcher) of our accelerator works in a pipelined fashion, the performance of ScalaBFS should scale linearly along the first direction (i.e., increasing the number of HBM PCs). We study the scalability of our accelerator in the first direction empirically by experiments conducted in Section \ref{sec:performance}, and focus our discussions of this section in the second direction (i.e., increasing PEs). By a model-based study, we intend to investigate the answer to the following question: \emph{Given a fixed number of HBM PCs, how many PEs in a PG should we choose to achieve the optimal performance?}

To simplify discussions, we consider a single HBM PC, on top of which a PG is constructed, and the number of PEs in the PG vary. We assume all processing units, i.e., PEs, the HBM reader, and the Vertex dispatcher, work in a pipelined fashion to mask delays in processing stages. As our PEs use double pump BRAMs as their local store on processing the bitmaps, each PE is capable of conducting two operations at each clock cycle. Therefore, when there are $N_{pe}$ PEs, we configure the data width of AXI bus (denoted as $DW$) in Equation \ref{equation:datawidth}. 

\begin{equation}
\begin{aligned}
DW = 2 \cdot N_{pe} \cdot S_v
\end{aligned}
\label{equation:datawidth}
\end{equation}

\noindent where $S_v$ denotes the storage size of a vertex. As each neighbor list read from an HBM PC is treated as a stream of vertices, which are classified and then sent to all participating PEs by the Vertex dispatcher, assuming each PE receives the same number of vertices during processing (perfect load balancing), to feed more PEs, we need longer AXI buses. 

Denote the PE's frequency as $F$, the bandwidth of a single HBM PC (denoted as $BW$) can be computed according to Equation \ref{equation:idealbandwidth}. 

\begin{equation}
\begin{aligned}
BW = 
&\begin{dcases}
DW \cdot F 
,&{DW \cdot F < BW_{MAX}}\\
BW_{MAX}
,& {DW \cdot F \ge BW_{MAX}}
\end{dcases}
\end{aligned}
\label{equation:idealbandwidth}
\end{equation}

\noindent where $BW_{MAX}$ denotes the maximum physical bandwidth of a single HBM PC. According to \cite{Wang:2020}, $BW_{MAX}$ is about 13.27GB/s. As HBM works in much higher frequency than PGs (e.g., the frequency of HBM in U280 is about 900 MHz, which is much higher than the FPGA circuits of ScalaBFS), we assume it produces a datum whose length is $DW$ for each cycle, but at the same time, $BW$ can not exceed its physical limit of $BW_{MAX}$. 

Now consider conducting a push-mode BFS iteration (pull-mode iterations exhibit similar patterns) in a given graph, a set of active vertices, each of which has a list of neighboring vertices stored in the edge array in Figure \ref{fig:partitionexamplegraph}, is going to be processed. We assume that each active vertex has $Len_{nl}$ neighboring vertices on average. Before retrieving the neighboring vertices, the HBM reader needs to read the offset values from the HBM. For each subgraph, assuming the length of data read from the offset array for each active vertex equals to the data width (i.e., $DW$), the percentage of HBM bandwidth paid on reading the neighbor lists $P_{nl}$ can be computed using Equation \ref{equation:percent}. 

\begin{equation}
\begin{aligned}
P_{nl} = \frac{Len_{nl} \cdot S_v}{DW + Len_{nl} \cdot S_v} 
\end{aligned}
\label{equation:percent}
\end{equation}

Accordingly, the portion of bandwidth paid on reading the neighbor lists can be computed using Equation \ref{equation:neighborlistbandwidth}.

\begin{equation}
\begin{aligned}
BW_{nl} = BW\cdot P_{nl} =
&\begin{dcases}
DW \cdot F \cdot P_{nl}
,&{DW \cdot F < BW_{MAX}}\\
BW_{MAX} \cdot P_{nl}
,& {DW \cdot F \ge BW_{MAX}}
\end{dcases}
\end{aligned}
\label{equation:neighborlistbandwidth}
\end{equation}

Regard each neighbor vertex as an edge connecting itself with a corresponding active vertex, by combining Equation \ref{equation:datawidth}, \ref{equation:percent} and \ref{equation:neighborlistbandwidth}, the theoretical performanceof a single PG (denoted as $Perf_{pg}.$, in number of traversed edges per second (TEPS)) can then be computed approximately in Equation \ref{equation:theoryperformance}. 

\begin{equation}
\begin{aligned}
&\mathrel{\phantom{=}} Perf_{pg}. \approx \frac{BW_{nl}}{S_v} = \\
&\begin{dcases}
\frac{2 N_{pe}\cdot F\cdot Len_{nl}}{2 N_{pe} + Len_{nl}}
,&{2 N_{pe} \cdot S_v \cdot F < BW_{MAX}}\\
\frac{BW_{MAX}\cdot Len_{nl}}{2 N_{pe}\cdot S_v+Len_{nl}\cdot S_v}
,& {2 N_{pe} \cdot S_v \cdot F \ge BW_{MAX}}
\end{dcases}
\end{aligned}
\label{equation:theoryperformance}
\end{equation}

Assuming that the vertex dispatcher is not the bottleneck (which means adding PC/PGs can improve the performance linearly, and our experiment can prove this), We can come up with our overall performance $Perf.$ in  Equation \ref{equation:overallperformance}, where $N_{pc}$ is the number of PC or PGs. 

\begin{equation}
\begin{aligned}
&\mathrel{\phantom{=}} Perf. = Perf_{pg} \cdot N_{pc}\\
&\begin{dcases}
\frac{2 N_{pe}\cdot F\cdot Len_{nl} \cdot N_{pc}}{2 N_{pe} + Len_{nl}}
,&{2 N_{pe} \cdot S_v \cdot F < BW_{MAX}}\\
\frac{BW_{MAX}\cdot Len_{nl} \cdot N_{pc}}{2 N_{pe}\cdot S_v+Len_{nl}\cdot S_v}
,& {2 N_{pe} \cdot S_v \cdot F \ge BW_{MAX}}
\end{dcases}
\end{aligned}
\label{equation:overallperformance}
\end{equation}

Also we can take the resource consumption into the consideration. For simplicity, only LUT usage is considered. Assuming each FIFO in our $k$ level vertex dispatcher costs $R_{FIFO}$ LUTs, each PE costs $R_{PE}$ LUTs, and the overall limit of LUTs is $R_{limit}$. Then we have the constraints as inequality \label{equation:inequality}. Notice that $N_{pe}$ must be power of 2 in our project.

\begin{equation}
\begin{aligned}
kN_{pe}^{\frac{1}{k}+1} \cdot R_{FIFO} + N_{PE} \cdot R_{PE} < R_{limit}
\end{aligned}
\label{equation:inequality}
\end{equation}

If we know $R_{limit}$, then we can figure out the maximum number of PEs we can have. For example on Xilinx Alveo U280, our maximum number of PE is 64.

\begin{figure}[t]
	\centering
	\setlength{\abovecaptionskip}{+5pt}
	\setlength{\belowcaptionskip}{-12pt}	
	\includegraphics[width = 3in]{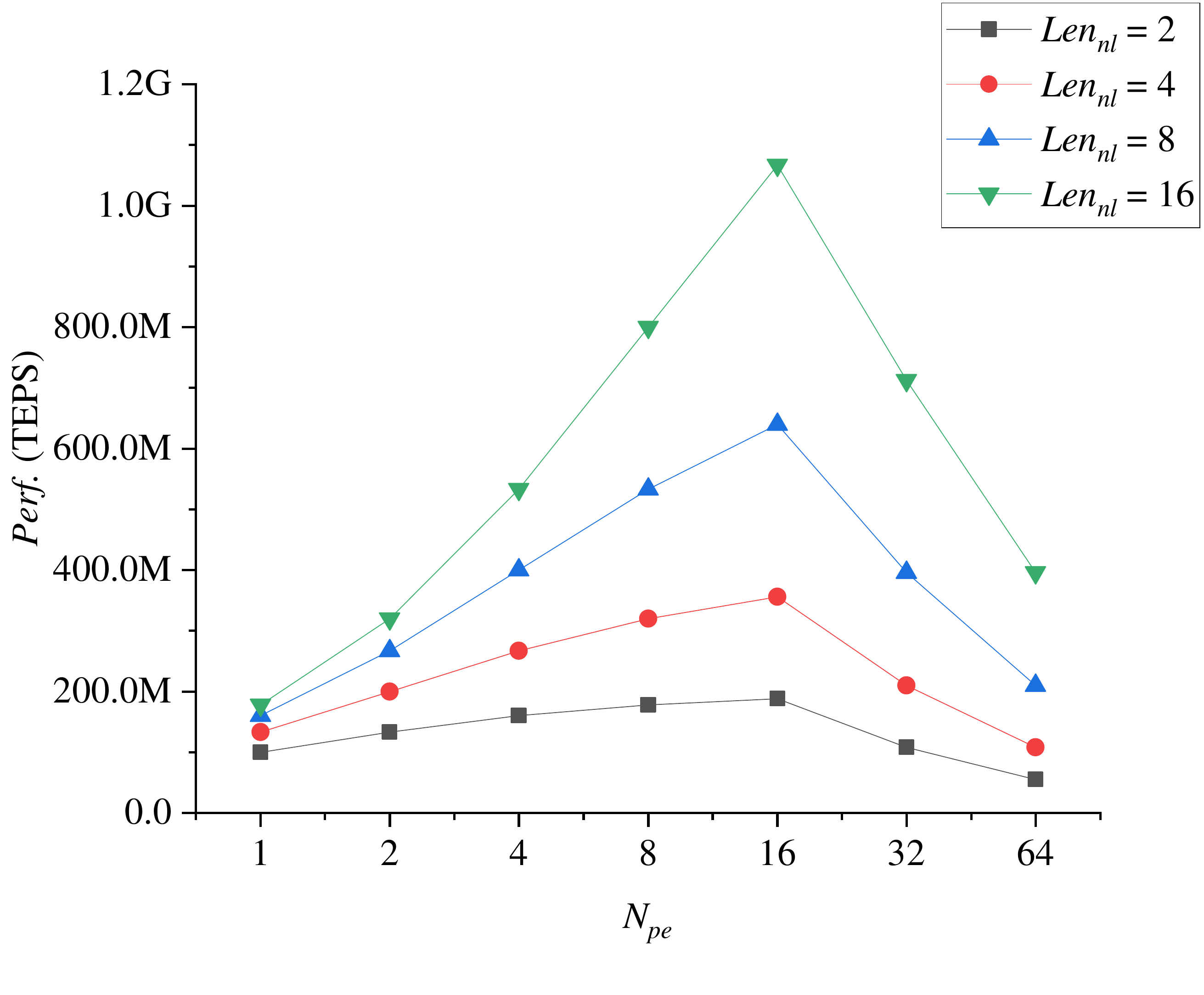}	
	\caption{Theoretical performance on a single HBM PC when $S_v = 32$ bits, $F = 100MHz$, $BW_{MAX} = 13.27GB/s$}
	\label{fig:theoryperformance}
\end{figure}

Let $S_v = 32$ bits, $F = 100MHz$ and $BW_{MAX} = 13.27GB/s$, we compute and illustrate $Perf.$ in Figure \ref{fig:theoryperformance}. From Figure \ref{fig:theoryperformance}, we can observe that: 1) with an equal number of PEs, the accelerator achieves better performance in graphs with larger $Len_{nl}$. 2) for a given graph (fixed $Len_{nl}$), increasing the number of PEs can improve the performance of the accelerator, and the performance gains are much higher for graphs with larger $Len_{nl}$s. Nevertheless, there is a break-point (i.e., 16 PEs), after which the performance will degrade when the number of PEs increases. 

The second observation is somewhat counter-intuitive since in the worst case, increasing the PEs should result in no detrimental effects on performance except for wasting logic resources. The underlying reason for such a phenomenon is that we configure the data width (i.e., $DW$) of the AXI bus according to the number of PEs. A larger number of PEs leads to larger $DW$ according to Equation \ref{equation:datawidth}, which further leads to smaller $P_{nl}$ in Equation \ref{equation:percent}, and consequently results in smaller $BW_{nl}$ and $Perf.$, if the HBM PC becomes saturated (i.e., the second conditions of Equation \ref{equation:neighborlistbandwidth} and \ref{equation:theoryperformance}). \emph{Such an observation reveals the fact that when given a fixed number of HBM PCs, there exists a configuration regarding the number of PEs for ScalaBFS to achieve its upper-bound performance}. We will further investigate this configuration by experiments in Section \ref{sec:performance} and verify the model of this section.

\section{System Evaluation}\label{sec:performance}

In this section, we will give the setups for our experiments in subsection \ref{subsec:experimentsetup}, present the resource consumption of ScalaBFS in subsection \ref{subsec:resourcesconsumption}, discuss hybrid-mode processing of the PEs in subsection \ref{subsec:hybridmodeprocessing}, address the scaling issues in subsection \ref{subsec:performancescaling}, study the HBM bandwidth utilization in subsection \ref{subsec:hbmbandwidth}, and compare the performances of ScalaBFS with other state-of-art systems in subsection \ref{subsec:comparewithothersystem}.

\subsection{Experiment Setups}\label{subsec:experimentsetup}

\noindent{\bf Hardware. }We run our experiments in a COTS PC server, which features two Xeon Silver 4110 CPU running at 2.10GHz, and a Xilinx Alveo U280 accelerator card \cite{XilinxU280:2019} attached to the PC server via PCIe bus. The U280 card has an HBM subsystem (shown in Figure \ref{fig:u280hbm}) of 8GB storage capacity. The HBM subsystem contains 32 pseudo channels, and provides a theoretical aggregated memory bandwidth up to 460GB/s. The Ultrascale+ FPGA in the U280 card contains 9.072MB BRAMs, 34.56MB URAMs and 1304K LUTs for implementing ScalaBFS. We implement ScalaBFS with Xilinx Vitis 2019.2. The host part of ScalaBFS is coded using OpenCL, and device part is programmed as an RTL kernel using Chisel language \cite{Bachrach:2012}. Thanks to the productivity and flexibility (with which we can easily vary the number of PCs and PEs in ScalaBFS by simply changing the parameters) of Chisel language, the RTL kernel design of ScalaBFS consists of only about 1700 lines of Chisel code. 

\noindent{\bf Workloads. }We choose four real-world graphs taken from \cite{Davis:2011} and ten synthetic RMAT graphs created using the Kronecker Generator (with parameters: A=0.57, B=0.19, C=0.19) from Graph 500 \cite{Richard:2010} as listed in Table \ref{tab:graphs} to evaluate the performance of ScalaBFS. In the names of RMAT graphs, the first number stands for the scale (i.e., the number of vertices) of the graph, and the second number stands for the average degree (i.e., dividing number of edges by the number of vertices). For example, ``RMAT18-8'' represents a synthetic graph with $2^{18}=256K$ vertices and $2^{18}\times8 = 2M$ edges. For an undirected graph, we convert each of its edges (except for the loop that connects the same vertex) into two directed edges with opposite directions.

On evaluating the performances, we use the notion of GTEPS (Giga Traversed Edges Per Second), which is computed by dividing the sum of outgoing or incoming neighbor list lengths of all visited vertices by the execution time of BFS. If an edge is ``visited'' more than once, it is counted only once. 

\begin{table}[t]
	\centering
	\setlength{\abovecaptionskip}{+5pt}
	\setlength{\belowcaptionskip}{-12pt}
	\caption{Graph datasets}
	\label{tab:graphs}
	\begin{tabular}{c|c|c|c|c}
		\hline
		\multirow{2}{*}{Graphs}
		& \#Vertices & \#Edges & Avg. &\multirow{2}{*}{Directed} \\
		&  (M) & (M) & Degree & \\
		\hline
		soc-Pokec (PK) &1.63 &30.62 & 18.75 & Y\\
		\hline
		soc-LiveJournal (LJ) & 4.85 & 68.99 & 14.23 & Y\\
		\hline
		com-Orkut (OR) & 3.07 & 234.37 & 76.28 & N \\
		\hline
		hollywood-2009 (HO) & 1.14 & 113.89 & 99.91 & N \\
		\hline
		\hline
		RMAT18-8 & 0.26 & 2.05 & 7.81 & N \\
		\hline
		RMAT18-16 & 0.26 & 4.03 & 15.39 & N \\
		\hline
		RMAT18-32 & 0.26 & 7.88 & 30.06 & N \\
		\hline
		RMAT18-64 & 0.26 & 15.22 & 58.07 & N \\
		\hline
		\hline
		RMAT22-16 & 4.19 & 65.97 & 15.73 & N \\
		\hline
		RMAT22-32 & 4.19 & 130.49 & 31.11 & N \\
		\hline
		RMAT22-64 & 4.19 & 256.62 & 61.18 & N \\
		\hline
		RMAT23-16 & 8.39 & 132.38 & 15.78 & N \\
		\hline
		RMAT23-32 & 8.39 & 262.33 & 31.27 & N \\
		\hline
		RMAT23-64 & 8.39 & 517.34 & 61.67 & N \\
		\hline
	\end{tabular}
\end{table}

\subsection{Resource Consumption}\label{subsec:resourcesconsumption}

\begin{table}[t]
	\centering
	\caption{Resource Utilization of ScalaBFS with typical configurations on U280 (VD stands for Vertex dispatcher)}
	\label{tab:resource}
	\begin{tabular}{|c|c|c|c|c|c|}
		\hline
		\#PC / \#PE & $f_{PE}$ / $f_{BRAM}$ &  \multirow{2}{*}{part} & \multirow{2}{*}{LUT} & \multirow{2}{*}{FF}  & \multirow{2}{*}{BRAM} \\
		& (MHz) & &  & & \\
		\hline
		\multirow{3}{*}{16 / 32} & \multirow{3}{*}{90 / 180} & total& 35.76\% & 10.29\% & 45.83\%  \\
		\cline{3-6}
		& & PGs & 7.68\%  & 0.76 \% & 35.71\%  \\
		\cline{3-6}
		& & VD & 16.71\%  & 1.17\% & 0\% \\
		\hline
		\multirow{3}{*}{32 / 32} & \multirow{3}{*}{90 / 180} & total& 39.93\% & 13.22\% & 46.33\%  \\
		\cline{3-6}
		& & PGs & 8.97\%  & 0.91\%& 35.71\%  \\
		\cline{3-6}
		& & VD & 16.66\%  & 1.17\%& 0\% \\
		\hline
		\multirow{3}{*}{32 / 64} & \multirow{3}{*}{90 / 180} & total& 42.08\% & 13.52\% & 48.21\%  \\
		\cline{3-6}
		& & PGs & 14.31\% & 1.50\%  & 38.10\%  \\
		\cline{3-6}
		& & VD & 13.40\% & 1.39\% & 0\% \\
		\hline
	\end{tabular}
\end{table}


Table \ref{tab:resource} lists the amounts of FPGA resources consumed by ScalaBFS in typical configurations (place and route results). When configured with 64 PEs and using all 32 PCs from HBM (i.e., the third configuration), the Vertex dispatcher of ScalaBFS uses a 3-layer crossbar, each layer of which consists of 16 4$\times$4 crossbars. For other configurations, the Vertex dispatcher uses full crossbars. 

Comparing the 16-PC/32-PE configuration with the 32-PC/32-PE configuration in Table \ref{tab:resource}, we can observe that the Vertex dispatchers of these two configurations consume almost the same amount of resources from U280. This is because both these two configurations of ScalaBFS use the same 32$\times$32 full crossbars with 32 PEs. There are two subgraphs stored in each HBM PC in the 16-PC/32-PE configuration, while there is only one subgraph stored in each HBM PC in the 32-PC/32-PE configuration. We can also observe that the PGs in the 32-PC/32-PE configuration consume more resources than those in the 16-PC and 32-PE configuration. This is because the 16-PC/32-PE configuration has only 16 PGs, each of which contains 2 PEs and an HBM reader (totally 16 HBM readers). In contrast, the 32-PC/32-PE configuration has 32 PGs, each of which contains one PE and also an HBM reader (totally 32 HBM readers). 

Comparing the 32-PC/32-PE configuration with the 32-PC/64-PE configuration in Table \ref{tab:resource}, we can observe that the Vertex dispatcher of the former configuration even consumes more resources than the latter configuration. This is because the 3-layer 4$\times$4 crossbars of the 32-PC/64-PE configuration consumes $3$ (layer) $\times 16$ (crossbars per layer) $\times 4 \times 4$ (FIFOs per crossbar) = 768 FIFOs, while the 32$\times$32 full crossbar of the 32-PC/32-PE configuration consumes $32^2 = 1024$ FIFOs. These numbers partially explain the differences in resource consumption of these two configurations.

From Table \ref{tab:resource}, we can observe that the PEs of ScalaBFS consume relatively small amounts of resources, even they are capable of hybrid (push-pull) mode processing. This is because the PEs reuse the circuits for push and pull modes, and the logic for memory accessing is decoupled from processing. Nevertheless, ScalaBFS stops at 64 PEs in the direction of scaling with more PEs in Table \ref{tab:resource}, since at our current stage of developing, higher numbers (e.g., 128) of PEs still lead to timing problems during the place and route phase. We will report performance results of ScalaBFS with more than 64 PEs in our future work. 

\subsection{Hybrid-mode Processing}\label{subsec:hybridmodeprocessing}

\begin{figure}[t]
	\setlength{\abovecaptionskip}{+5pt}
	\setlength{\belowcaptionskip}{-12pt}
	\centering
	\includegraphics[width = 3.1in]{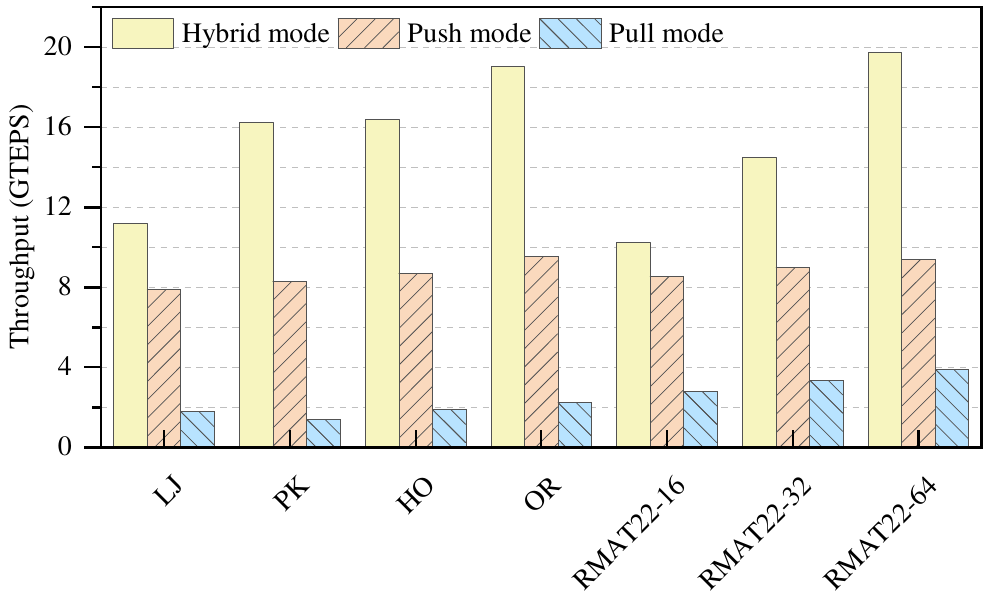}	
	\caption{Performances of ScalaBFS configured with 32 HBM PCs and 64 PEs, PEs work in different processing modes.}
	\label{fig:hybrid}
\end{figure}
In this section, We leverage the 32-PC/64-PE configuration that yields the highest GTEPS to examine the effect of hybrid-model processing, compared with pull and push models, where two PEs are associated with a PC. Figure \ref{fig:hybrid} illustrates the absolute throughput, in terms of GTEPS, of three models. We have two observations. First, the hybrid mode leads to 1.20$\sim$2.10x (or 3.65$\sim$11.52x) throughput improvement over the push (or pull) model, Second, ScalaBFS is able to achieve higher performance improvement when processing denser graphs, as the BFS processing in the hybrid mode effectively reduces the number of unnecessary memory accesses to the neighbor lists, coincident with previous works \cite{Beamer:2012}, \cite{Umuroglu:2015} and \cite{Zhang:2018}. For example,  when processing the dense RMAT22-64 graph, ScalaBFS achieves its peak performance of 19.7 GTEPS. Therefore, we let PEs of ScalaBFS work in the hybrid mode in the following experiments. 


\subsection{Performance Scaling}\label{subsec:performancescaling}

For ScalaBFS, there are two scaling directions: increasing HBM PCs and increasing PEs. We first study the performance scaling of ScalaBFS by increasing the number of HBM PCs, and then study the performance scaling by increasing the number of PEs. 

\begin{figure}[t]
	\setlength{\abovecaptionskip}{+5pt}
	\setlength{\belowcaptionskip}{-12pt}
	\centering
	\includegraphics[width = 3in]{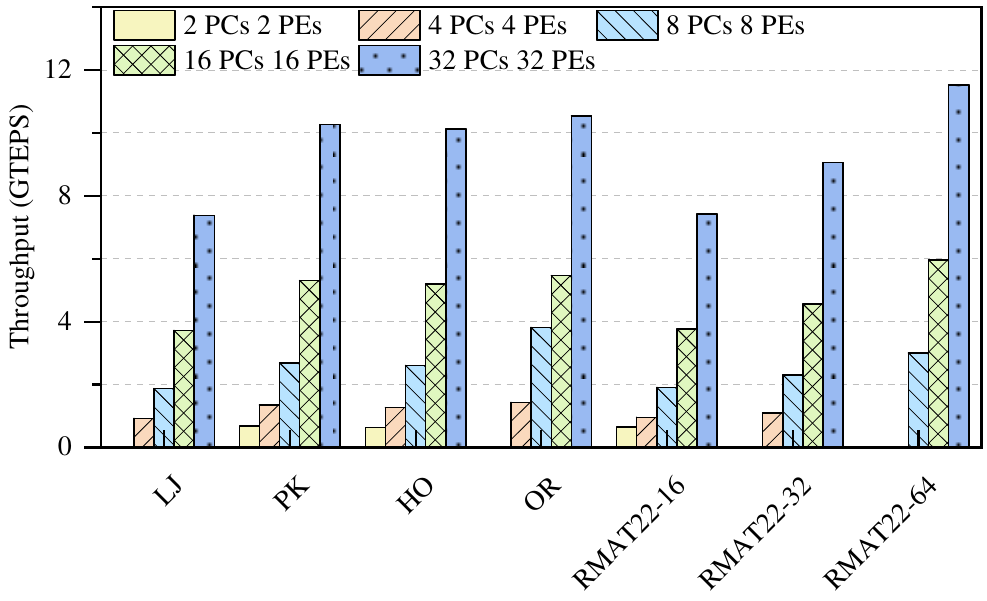}	
	\caption{Performances of ScalaBFS using increasing numbers of HBM PCs.}
	\label{fig:morePCs}
\end{figure}

\noindent{\bf Effect of HBM PC. }We examine the effect of HBM PC with the configuration of a single PE for each PG (thus on each HBM PC). Figure \ref{fig:morePCs} illustrates the throughput improvement with a increasing number of HBM PCs. We observe that ScalaBFS achieves almost linear speedup with respect to the number of HBM PCs, indicating that more the decoupled design of ScaleBFS that allows decoupling memory accessing from processing leads to high efficency and effectiveness when performing BFS. 

\begin{figure}[t]
	\setlength{\abovecaptionskip}{+5pt}
	\setlength{\belowcaptionskip}{-12pt}
	\centering
	\includegraphics[width = 3.1in]{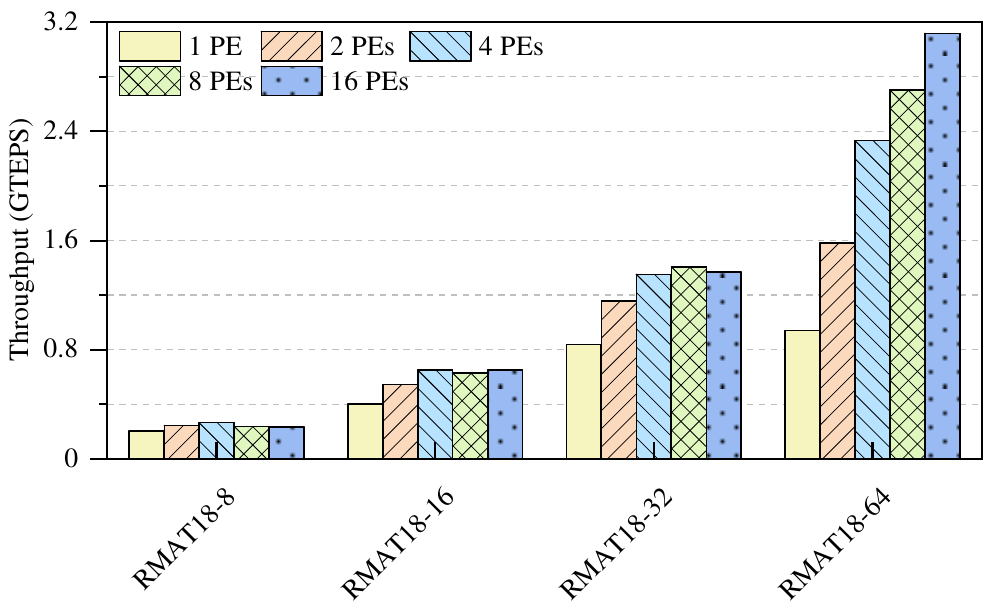}\
	\caption{Performances with different numbers of PEs within a single HBM PC}
	\label{fig:morePEs}
\end{figure}

\noindent{\bf Effect of PE. }We examine the effect of PE within a PG (i.e., on a HBM PC). Since a single HBM PC provides only 2 Gbits storage capacity that limits the size of the graph data, we use small synthetic graphs with a scale of $2^{18}$ vertices, as illustrated in Figure \ref{fig:morePEs}. We observe that more PEs lead to higher performance of ScalaBFS, especially when the vertices of the input graph have higher average degrees. 

However, the performance scaling stops at specific break-points. For example, the performances for the RMAT18-8 and RMAT18-16 graph stop at 4 PEs, that for the RMAT18-32 graph stops at 8PEs. Comparing Figure \ref{fig:morePEs} and Figure \ref{fig:theoryperformance}, we can observe that the trends are similar, except that the break-points show up earlier for graphs whose vertices have smaller numbers of neighboring vertices in real systems than in theory. This is because in the theoretical study in Section \ref{sec:performancemodel}, we assume the load-balancing status is perfect, i.e., the Vertex dispatcher evenly distributes vertices among the PEs. Nevertheless, such an assumption may be compromised by the structure of the input graphs in real systems. Whereas, experimental results from Figure \ref{fig:morePEs} still suggests that the optimal configuration on the number of PEs configured to each HBM PC in ScalaBFS is about 4 to 8 for sparse graphs, and it is about 8 to 16 for dense graphs. By this observation, \emph{our 32-PC/64-PE configuration listed in Table \ref{tab:resource} has not fully exploited the HBM subsystem of U280, since each HBM PC is configured with only two PEs in such configuration}. Deductively, we should configure about 128 (32$\times$4) PEs or 512 (32$\times$16) PEs for ScalaBFS to achieve the optimal performance in the sparse or dense graph respectively on U280.

Comparing Figure \ref{fig:morePEs} with Figure \ref{fig:morePCs}, we can observe that the performance gain from increasing the number of PEs is much less than from increasing the number of HBM PCs. For example, increasing the number of PEs from one to two or one to four leads to about 1.68x or 2.48x speedup, even on the (dense) RMAT18-64 graph. \emph{Such an observation suggests that ScalaBFS should prioritize the scaling in the direction of using more HBM PCs, rather than increasing the number of PEs, especially when the amount of FPGA logic resources is limited.} It is also the reason why the 32-PC/64-PE configuration of ScalaBFS can achieve the best performance so far on U280. 

\subsection{HBM Bandwidth Utilization}\label{subsec:hbmbandwidth}

\begin{figure}[t]
	\setlength{\abovecaptionskip}{+5pt}
	\setlength{\belowcaptionskip}{-12pt}
	\centering
	\includegraphics[width = 3.5in]{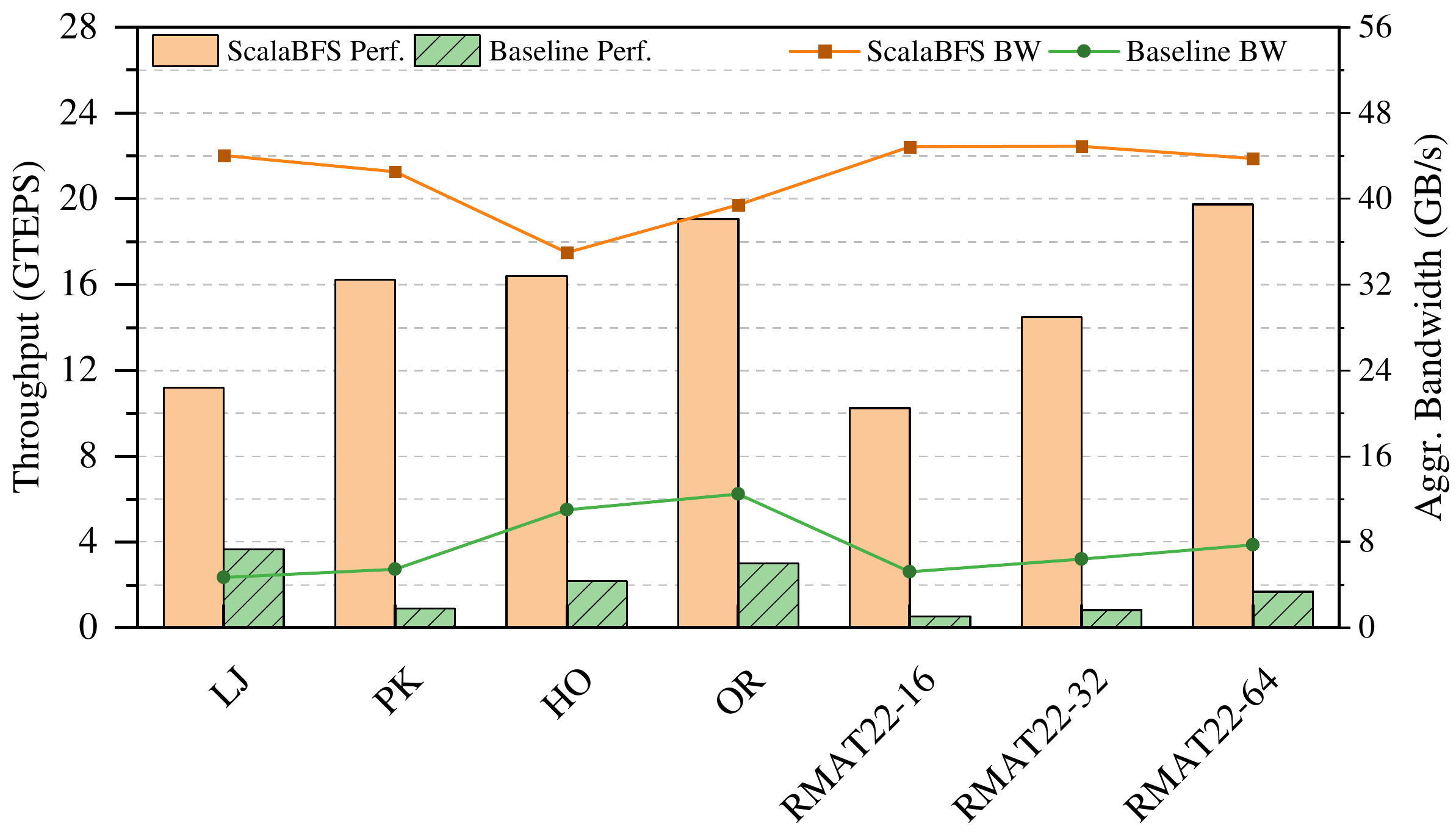}\
	\caption{Performances and aggregated bandwidths of ScalaBFS (with 32 HBM PCs and 64 PEs) and baseline case}
	\label{fig:BW_compare}
\end{figure}

We measure the aggregated bandwidth (summation of bandwidths of 32 PCs) of HBM during the execution of BFS on various graph datasets with ScalaBFS. For comparison, we introduce the \emph{baseline} case. Contrary to ScalaBFS where the edge data of an input graph are partitioned (shown in Figure \ref{subfig:patitioned}) to evenly distributed to all HBM PCs, in the baseline case, the edge data are \emph{not} partitioned (shown in Figure \ref{subfig:csrcsc}), and placed in the HBM PCs sequentially (starting from PC0). For the data distribution, the HBM readers of the PGs in ScalaBFS only need to access its corresponding PC, while an HBM reader in the baseline case have to cross (possibly) multiple PCs to access the adjacency lists of the vertices assigned to its PG.

Figure \ref{fig:BW_compare} reports the aggregated bandwidths of HBM, as well as the performances of both ScalaBFS and the baseline case, when both systems are employed to process the same graphs. From Figure \ref{fig:BW_compare}, we can observe that the baseline cases poorly use the HBM (small aggregated bandwidths) and consequently produce poor performances on BFS. The reasons for the small aggregated bandwidths in these baseline cases are that: 1) the PGs have to read the required edge data from remote PCs during computation, which places pressure on the switch network of HBM in Figure \ref{fig:u280hbm}, and further leads to poor bandwidths as in Figure \ref{fig_effect_mini_switch}; 2) the edge data of our chosen graphs are relatively small, and are thus stored in the PCs with small suffixes. During execution, such data placement scheme leads to unbalanced accesses and further limits the achievable bandwidths of HBM. 

Compared with the baseline cases, ScalaBFS achieves much higher aggregated bandwidths by evenly distributing the edge data in HBM PCs, and at the same time, confining the PGs to access their corresponding PCs. From Figure \ref{fig:BW_compare}, we can observe that the maximal achieved bandwidth is about 46GB/s. Since the RTL of ScalaBFS runs at 90MHz, the burst length is 128bits (2 PEs to each PC), the theoretical upper-bound aggregated bandwidth is about: 90MHz $\times$ 128/8 Byte $\times$ 32 PCs $\approx$ 46.08GB/s (i.e., the measured bandwidth is close to the upper-bound in theory). Nevertheless, such bandwidth is still much smaller than the theoretical bandwidth (up to 460GB/s) of HBM in U280. The reasons are that: 1) the access pattern of BFS is random and irregular, which limits the achievable bandwidths of each HBM PC using DRAM; 2) the relatively low clock frequency (i.e., 90MHz) limits the maximum achievable bandwidth as computed above. 
\subsection{Comparison with Other Existing Systems}\label{subsec:comparewithothersystem}

\begin{figure}[t]
	\setlength{\abovecaptionskip}{+5pt}
	\centering
	\includegraphics[width = 3.1in]{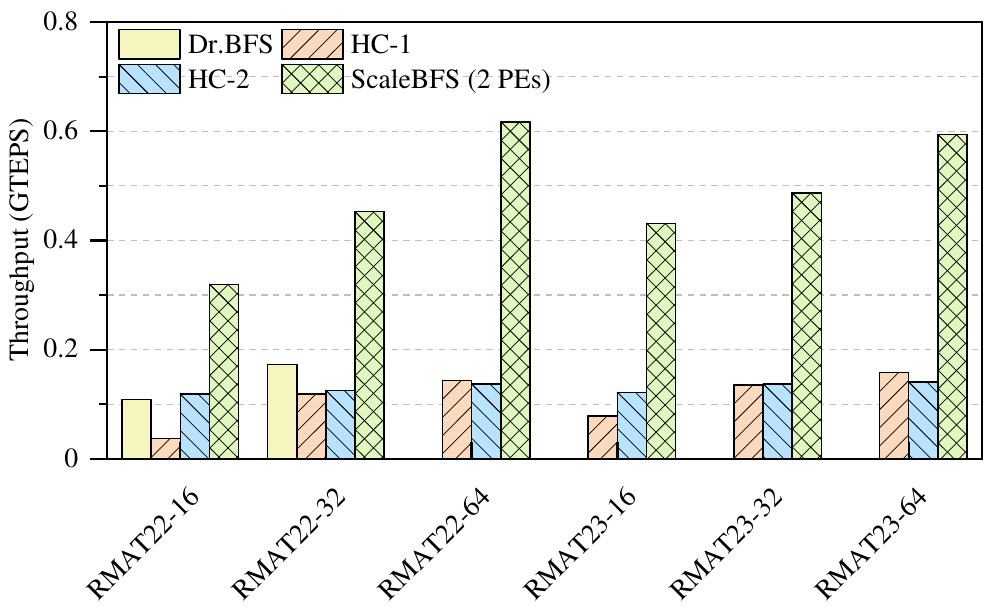}\
	\caption{Average single DRAM channel throughput in ScalaBFS and other FPGA-based BFS accelerators}
	\label{fig:Perf_compare_other}
\end{figure}

\noindent{\bf Compared with Other FPGA-based Systems. }In the existing works regarding FPGA-based BFS accelerators, \cite{Betkaoui:2012} and \cite{Attia:2014} achieve 2.5 GTEPS by using 16xDDR2 on the Convey machines and Dr. BFS \cite{Finnerty:2019} that achieves about 470 MTEPS by using 2xDDR4, ScalaBFS achieves its peak performance of 19.7 GTEPS and yields about 7.9x over \cite{Betkaoui:2012} and \cite{Attia:2014}, since ScalaBFS exploits HBM's high memory bandwidth and massive FPGA resources for PE-level parallelism. AS different accelerators use different FPGA platforms with different numbers of DRAM channels, to be fair, Figure \ref{fig:Perf_compare_other} compares the BFS performance of ScalaBFS and other similar systems using a \emph{single} DRAM channel. From Figure \ref{fig:Perf_compare_other}, we can observe that ScalaBFS is much faster than existing systems even in the context of single memory channel performance, which further explains the high BFS performance speedups over existing systems. 

Note that the performance of ScalaBFS on U280 is still far from the theoretical performance of 45.8 GTEPS \cite{Zhang:2018} on bitmap operations (key operations of BFS) using the HMC device that supports the processing-in-memory (PIM) technology. Nevertheless, we should notice that the above experiments are conducted on the real FPGA board U280, which limits the performance of ScalaBFS with its fixed amount of physical resources, e.g., the number of HBM PCs and LUTs. We believe with the technology progresses, ScalaBFS will continuously achieve higher performance on future FPGA cards, that feature more HBM stacks and more logic resources, with its scalability.

\noindent{\bf Compared with GPU-based Systems. }We further compare the performances of ScalaBFS on U280 with those of Gunrock \cite{Wang:2017}, which is a popular accelerator for graph processing on GPUs, running on Nvidia V100 GPU (model \texttt{SXM2}, configured with 640 tensor cores and 5120 CUDA cores running at 1.53 GHz, 4 HBM2 stacks (totally 64 PCs, 16 GB storage space), consuming 300 watts \cite{TeslaV100datasheet:2019}), when conducting BFS in real-world graphs in Table \ref{tab:comparewithGunrock}. Gunrock also adopts the hybrid processing mode as in ScalaBFS when conducting BFS. We fine-tune the parameters of Gunrock during the experiments such that Gunrock achieves its best performance on V100. From Table \ref{tab:comparewithGunrock}, we can observe that for sparse real-world graphs (i.e., \texttt{PK} and \texttt{LJ}) whose vertices have short neighbor lists, the performances of ScalaBFS are very close to those of Gunrock. This is because, when conducting BFS in these sparse graphs, the memory accesses towards the HBM are of smaller burst lengths, which makes the HBM to be the bottleneck of the system. ScalaBFS uses Algorithm \ref{alg:bfs-optimize} where the bitmaps stored in the BRAM absorb large amount of memory accesses to conduct BFS, and thus achieves equivalent performances by using only 32 HBM PCs to those achieved in Gunrock using 64 HBM PCs. Moreover, the power efficiencies of ScalaBFS on U280 are about 5.68$\sim$10.19x better than those of Gunrock on V100, due to the low power consumption of U280. We read the on-board power meter by using the Xilinx Board Utility (xbutil) \cite{xbutil:2020}, which reports 32 watts during the processing of all graphs in Table \ref{tab:comparewithGunrock}.

\begin{table}[t]
	\centering
	\setlength{\belowcaptionskip}{-12pt}	
	\caption{Performance comparison between GunRock and ScalaBFS (32-PC/64-PE configuration)}
	\label{tab:comparewithGunrock}
	\resizebox{\columnwidth}{!}{%
		\begin{tabular}{|c|c|c|c|c|}
			\hline
			& \multicolumn{2}{c|}{Gunrock on V100} & \multicolumn{2}{c|}{ScalaBFS on U280} \\ \hline
			Datasets &
			\begin{tabular}[c]{@{}c@{}}Throughput\\ (GTEPS)\end{tabular} &
			\begin{tabular}[c]{@{}c@{}}Power eff.\\ (GTEPS/watt)\end{tabular} &
			\begin{tabular}[c]{@{}c@{}}Throughput\\ (GTEPS)\end{tabular} &
			\begin{tabular}[c]{@{}c@{}}Power eff.\\ (GTEPS/watt)\end{tabular} \\ \hline
			PK       & 14.9          & 0.050        & 16.2          & 0.506         \\ \hline
			LJ & 18.5          & 0.062        & 11.2          & 0.350         \\ \hline
			OR       & 150.6         & 0.502        & 19.1          & 0.597         \\ \hline
			HO  & 73            & 0.243        & 16.4          & 0.513         \\ \hline
		\end{tabular}%
	}
\end{table}

From Table \ref{tab:comparewithGunrock}, we can also observe that for dense real-world graphs (i.e., \texttt{OR} and \texttt{HO}) whose vertices have long neighbor lists, the performances of ScalaBFS are only about 0.13$\sim$0.22x of the performances of Gunrock. The reason is that the dense graphs lead to memory accesses with much larger burst lengths during processing, which makes the processing elements to be the real bottlenecks of the system. To this end, Gunrock enjoys the huge memory bandwidth provided by the 64 PCs of the V100's HBM and takes advantage of the large number of high-frequency hardware cores of the V100 to win the battle. Such a performance disadvantage of ScalaBFS also suggests that when conducting BFS in dense graphs, we need to build enough processing elements in an FPGA-based BFS accelerator to achieve comparable or even higher performance than GPUs. 

\section{Conclusions and Future Works}\label{sec:conclusion}

In this paper, we propose ScalaBFS, an open-source BFS accelerator built on the FPGA-HBM platform. ScalaBFS features hybrid-mode processing elements that sufficiently exploit the memory bandwidth of HBM, and decoupled circuits for processing and memory accessing that scales its performance with the increasing HBM PCs. By fully using the 32 HBM pseudo channels (PCs), the performance of ScalaBFS arrives at 19.7 GTEPS when conducting BFS in real-world and synthetic scale-free graphs on Xilinx Alveo U280 Data Center Accelerator card. The future works of ScalaBFS include further fine-tuning of the system, supporting more processing elements on U280, processing larger graphs, and extending itself to a general graph-processing framework that is capable of conducting other graph algorithms.


\bibliographystyle{IEEEtran}
\bibliography{fpga-bib}


\end{document}